\documentclass{jfm}
\usepackage{graphicx}
\usepackage{natbib}
\ifCUPmtlplainloaded \else
  \checkfont{eurm10}
  \iffontfound
    \IfFileExists{upmath.sty}
      {\typeout{^^JFound AMS Euler Roman fonts on the system,
                   using the 'upmath' package.^^J}%
       \usepackage{upmath}}
      {\typeout{^^JFound AMS Euler Roman fonts on the system, but you
                   dont seem to have the}%
       \typeout{'upmath' package installed. JFM.cls can take advantage
                 of these fonts,^^Jif you use 'upmath' package.^^J}%
      }
  \else
  \fi
\fi
\ifCUPmtlplainloaded \else
  \checkfont{msam10}
  \iffontfound
    \IfFileExists{amssymb.sty}
      {\typeout{^^JFound AMS Symbol fonts on the system, using the
                'amssymb' package.^^J}%
       \usepackage{amssymb}%
       \let\le=\leqslant  
       \let\ge=\geqslant  
      }{}
  \fi
\fi
\ifCUPmtlplainloaded \else
  \IfFileExists{amsbsy.sty}
    {\typeout{^^JFound the 'amsbsy' package on the system, using it.^^J}%
     \usepackage{amsbsy}}
    {\providecommand\boldsymbol[1]{\mbox{\boldmath $##1$}}}
\fi
\def\be{\begin{equation}}
\def\ee{\end{equation}}
\def\ba{\begin{eqnarray}}
\def\ea{\end{eqnarray}}
\def\pa{\parallel}
\def\kpa{k_{\parallel}}
\def\kpn{k_{\perp}}

\def\bbl{{\bf b_0}}
\providecommand\Ome{\boldsymbol{\Omega_0}}
\def\Omes{\Omega_0}
\def\uu{{\bf u}}
\def\bb{{\bf b}}
\def\ww{{\bf w}}
\def\jj{{\bf j}}
\def\kk{{\bf k}}
\def\pp{{\bf p}}
\def\qq{{\bf q}}
\def\ppa{p_{\parallel}}
\def\qpa{q_{\parallel}}
\def\ppn{p_{\perp}}
\def\qpn{q_{\perp}}
\def\diso{d\ppn d\qpn d\ppa d\qpa}
\def\nnn{{\tilde n}}
\def\mmm{{\tilde m}}

\def\kkpr{{\bf k}^{\prime}}
\def\ep{{\bf \hat{e}}_{\parallel}}
\def\ek{{\bf \hat{e}}_k}
\def\eep{{\bf \hat{e}}_p}

\def\nn{\bf \hat{n}}
\def\ete{{\bf \hat{e}}_{\theta}}
\def\efi{{\bf \hat{e}}_{\Phi}}
\def\kkp{{\bf k}_{\perp}}
\def\ols{\omega_{\Lambda}^s}
\def\olsp{\omega_{\Lambda_p}^{s_p}}
\def\olsq{\omega_{\Lambda_q}^{s_q}}
\def\hh{{\bf h^{\Lambda}_k}}
\def\Lp{\Lambda^{\prime}}

\def\aak{a_{\Lambda}^s}
\def\aap{a_{\Lambda_p}^{s_p}}
\def\aaq{a_{\Lambda_q}^{s_q}}
\def\aakp{a_{\Lambda^{\prime}}^{s^{\prime}}}
\def\aakdp{a_{\Lambda^{\prime \prime}}^{s^{\prime \prime}}}
\def\qls{q_{\Lambda}^s}
\def\qlsprime{q_{\Lambda^{\prime}}^{s^{\prime}}}
\def\qlsdprime{q_{\Lambda^{\prime \prime}}^{s^{\prime \prime}}}
\def\qlsp{q_{\Lambda_p}^{s_p}}
\def\qlsq{q_{\Lambda_q}^{s_q}}
\def\alf{\xi_{\Lambda}^s}
\def\alfp{\xi_{\Lambda_p}^{s_p}}
\def\alfq{\xi_{\Lambda_q}^{s_q}}
\def\alfm{\xi_{\Lambda}^{-s}}
\def\alfmp{\xi_{\Lambda_p}^{-s_p}}
\def\alfmq{\xi_{\Lambda_q}^{-s_q}}
\def\alfprime{\xi_{\Lambda^{\prime}}^{s^{\prime}}}
\def\alfmprime{\xi_{\Lambda^{\prime}}^{-s^{\prime}}}
\def\alfdprime{\xi_{\Lambda^{\prime \prime}}^{s^{\prime \prime}}}
\def\alfmdprime{\xi_{\Lambda^{\prime \prime}}^{-s^{\prime \prime}}}
\def\ie{{\it i.e.}\ }

\newsavebox{\astrutbox}
\sbox{\astrutbox}{\rule[-5pt]{0pt}{20pt}}

\newcommand\eg{e.g.\ }

\title[Weak turbulence theory for rotating MHD and planetary dynamos]
{Weak turbulence theory for rotating magnetohydrodynamics and planetary dynamos}
\author[S\'ebastien Galtier]{S\'ebastien Galtier
\thanks{Email address for correspondence: sebastien.galtier@lpp.polytechnique.fr}}
\affiliation{Laboratoire de Physique des Plasmas, \'Ecole Polytechnique, F-91128 Palaiseau Cedex, France}

\pubyear{2014}
\volume{999}
\pagerange{100--133}

\date{\today; revised ?; accepted ?.}

\begin{document}

\maketitle

\begin{abstract}
A weak turbulence theory is derived for magnetohydrodynamics under rapid rotation and in the presence of a large-scale 
magnetic field. The angular velocity $\Ome$ is assumed to be uniform and parallel to the constant Alfv\'en speed $\bbl$. 
Such a system exhibits left and right circularly polarized waves which can be obtained by introducing the magneto-inertial 
length $d \equiv b_0/\Omes$. In the large-scale limit ($kd \to 0$; $k$ being the wave number), the left- and right-handed 
waves tend respectively to the inertial and magnetostrophic waves whereas in the small-scale limit ($kd \to + \infty$) pure 
Alfv\'en waves are recovered. By using a complex helicity decomposition, the asymptotic weak turbulence equations are 
derived which describe the long-time behavior of weakly dispersive interacting waves {\it via} three-wave interaction processes. 
It is shown that the nonlinear dynamics is mainly anisotropic with a stronger transfer perpendicular ($\perp$) than parallel 
($\pa$) to the rotating axis. The general theory may converge to pure weak inertial/magnetostrophic or Alfv\'en wave turbulence 
when the large or small-scales limits are taken respectively. Inertial wave turbulence is asymptotically dominated by the kinetic
energy/helicity whereas the magnetostrophic wave turbulence is dominated by the magnetic energy/helicity. For both regimes 
a family of exact solutions are found for the spectra which do not correspond necessarily to a maximal helicity state. 
It is shown that the hybrid helicity exhibits a cascade whose direction may vary according to the scale $k_f$ at which the 
helicity flux is injected with an inverse cascade if $k_fd < 1$ and a direct cascade otherwise. 
The theory is relevant for the magnetostrophic dynamo whose main applications are the Earth and giant planets for which 
a small ($\sim 10^{-6}$) Rossby number is expected.
\end{abstract}


\section{Introduction}
Rotation is a commonly observed phenomenon in astronomy: planets, stars and galaxies all spin around their axis. 
The rotation rate of planets in the solar system was first measured by tracking visual features whereas stellar rotation is 
generally measured through Doppler shift or by following the magnetic activity. One consequence of the Sun rotation is 
the formation of the Parker interplanetary magnetic field spiral well detected by space crafts, whereas the Earth rotation 
has a strong impact on the turbulent dynamics of large-scale geophysical flows. These few examples show that the study 
of rotating flows interests a wide range of problems, ranging from engineering (turbomachinery) to geophysics (oceans, earth's 
atmosphere, gaseous planets), weather prediction and turbulence \citep{davidsonbook}. 
Rotation is often coupled with other dynamical factors, it is therefore important to isolate the effect of the Coriolis force to 
understand precisely its impact. The importance of rotation can be measured with the Rossby number:
\be
Ro = {U_0 \over L_0 \Omega_0} \, ,
\ee 
where $U_0$, $L_0$ and $\Omega_0$ are respectively typical velocity, length-scale and rotation rate. This dimensionless 
number measures the ratio of the advection term on the Coriolis force in the Navier-Stokes equations, also a small value of
$Ro$ means a dynamics mainly driven by rotation. Typical large-scale planetary flows are characterized by $Ro \sim 0.1$ 
\citep{shirley} whereas the liquid metals (mainly iron) in the Earth's outer core are much more affected by rotation with 
$Ro \sim 10^{-6}$ \citep{Roberts}. Note that for a giant planet like Jupiter in which liquid metallic hydrogen is present in 
most of the volume, it is believed that the Rossby number may even be smaller \citep[see \eg][]{jones}. These situations 
contrast with the solar convective region where the magnetic field is believed to be magnified and for which $Ro \sim 1$. 

Inertial waves are a ubiquitous feature of neutral fluids under rapid rotation \citep{greenspan}. Although much is known about 
their initial excitation, their nonlinear interactions is still a subject of intense research. 
Many papers have been devoted to pure rotating turbulence ($R_o \le1$) but because of the different nature of the investigations 
(theoretical, numerical and experimental) it is difficult in general to compare directly the results obtained. From a theoretical point of 
view it is convenient to use a spectral description in terms of continuous wave vectors with the unbounded homogeneity assumption 
in order to derive the governing equations for the energy, kinetic helicity and polarization spectra \citep{Cambon89}. Although such 
equations introduce transfer terms which remain to be evaluated consistently, it is already possible to show with a weakly nonlinear 
resonant waves analysis \citep{waleffe2} the anisotropic nature of that turbulence with a nonlinear transfer preferentially in the 
perpendicular (to $\Ome=\Omega {\bf e_\parallel}$) direction. For moderate Rossby numbers the eddy damped quasi-normal 
Markovian model may be used as a closure \citep{Cambon97}, whereas in the small Rossby number limit the asymptotic weak 
turbulence theory can be derived rigorously \citep{galtier03a}. In the latter case, it was shown that the wave modes ($\kpa>0$) are 
decoupled from the slow mode ($\kpa=0$) which is not accessible by the theory, and the positive energy flux spectra were also 
obtained as exact power law solutions. The weak turbulence regime was also investigated numerically and it was shown in particular 
that the two-dimensional manifolds is an integrable singularity at $k_\parallel=0$, which is related to the scaling of the energy spectrum 
$\propto k_\parallel^{-1/2}$, and that the energy cascade goes forward \citep{Bellet}. Recently, the problem of confinement has 
been addressed explicitly in the inertial wave turbulence theory using discrete wave numbers \citep{scott}: three asymptotically distinct 
stages in the evolution of the turbulence are found with finally a regime dominated by resonant interactions.
Pseudo-spectral codes are often used to investigate numerically homogeneous rotating turbulence \citep[see \eg][]{mininni10a}. 
Several questions have been investigated like the origin of the anisotropy or of the inverse cascade observed when a forcing is applied at 
intermediate scale $k_f$. However, according to the question addressed the results may be affected by the discretization and by finite-box 
effects at too small Rossby number and too long elapsed time \citep{smith2,bourouiba08}. This seems to be the case in 
particular for the question of the inverse cascade mediated by the decoupling of the slow mode. 
For example, it was found that the one-dimensional isotropic energy spectrum $E(k) \sim k^{-x}$ may follow two different power laws with 
$2 \le x \le 2.5$ at small-scale ($k>k_f$) and $x \simeq 3$ at large-scale ($k<k_f$) \citep{smith1}. But it was also shown that the scaling at 
large-scale was strongly influenced by the value of the aspect ratio between the parallel and the perpendicular (to $\Ome$) resolution, a 
small aspect leading to a reduction of the number of available resonant triads, hence an alteration of the spectrum with the restoration 
of a $k^{-5/3}$ spectrum for small enough vertical resolution.
Several experiments have been devoted to rotating turbulence with different types of apparatus 
\citep{hop,jacquin,baroud,morize,vanbokhoven}. Contrary to the theory and the simulation, it is very challenging to reproduce 
experimentally the conditions of homogeneous turbulence. Nevertheless, one of the main results reported is that the rotation leads 
to a bi-dimensionalisation of an initial homogeneous isotropic turbulence with anisotropic spectra where energy is preferentially 
accumulated in the perpendicular (to $\Ome$) wave numbers $k_\perp$. Energy spectra with $x \ge 2$ were experimentally observed 
\citep{baroud,morize,vanbokhoven} revealing a significant discrepancy with the isotropic Kolmogorov spectrum ($x=5/3$) for non-rotating 
fluids. Note that the wave number entering in the spectral measurements corresponds mainly to $k_\perp$. Recently, direct measurements 
of energy transfer have been made in the physical space by using third-order structure function \citep{lamriben} and an increase of 
anisotropy at small scales has been found in agreement with some theoretical studies \citep{jacquin,galtier03a,galtier09r,Bellet}. 
The role of kinetic helicity -- which quantifies departures from mirror symmetry \citep{moffatt} -- on rotating fluids has been the subject 
of few studies. One reason is that it is difficult to measure the helicity production from experimental studies. The other reason is 
probably linked to the negligible effect of helicity on energy in non-rotating turbulence. Indeed, in this case one observes a joint 
constant flux cascade of energy and helicity with a $k^{-5/3}$ spectrum for both quantities \citep{chen1,chen2}. But recently, several 
numerical simulations have demonstrated the surprising strong impact of helicity on fast rotating hydrodynamic turbulence 
\citep{mininni10a,mininni10b,teitelbaum09,mininni12} 
whose main properties can be summarized as follows. When the (large-scale) forcing applied to the system injects only negligible 
helicity, the dynamics is mainly governed by a direct energy cascade compatible with an energy spectrum $E(\kpn) \sim \kpn^{-5/2}$ 
which is precisely the weak turbulence prediction \citep{galtier03a}. However, when the helicity injection becomes so important that the 
dynamics is mainly governed by a direct helicity cascade, different scalings are found following the empirical law:
\be
n+\nnn = -4 \, ,
\label{nnnn}
\ee 
where $n$ and $\tilde n$ are respectively the power law indices of the one-dimensional energy and helicity spectra. This law 
cannot be explained by a consistent phenomenology where anisotropy is used which renders the relation (\ref{nnnn}) highly 
non-trivial. As shown by \cite{galtier2014}, an explanation can only be found when a rigorous analysis is made on the weak 
turbulence equations: the relation corresponds in fact to the finite helicity flux spectra which are exact solutions of the equations. 

It has been long recognized that the Earth's magnetic field is not steady \citep{Finlay10}. Changes occur across a wide range
of timescales from second -- because of the interactions between the solar wind and the magnetosphere -- to several tens of 
millions years which is the longest timespan between polarity reversals. To understand the generation and the maintain of 
a large-scale magnetic field, the most promising mechanism is the dynamo \citep{pouquet76,moffatt78,brandenburg}. Dynamo is 
an active area of research where dramatic developments have been made in the past several years \citep{dormy00}. The subject 
concerns primary the Earth where a large amount of data is available which allows us to follow \eg the geomagnetic polarity 
reversal occurrences over million years \citep{finlay03,Roberts}. This chaotic behavior contrasts drastically with the surprisingly 
regularity of the Sun which changes its magnetic field lines polarity every $\sim 11$ years. It is believed that the three main 
ingredients for the geodynamo problem are the Coriolis, Lorentz-Laplace and buoyancy forces. The latter force may be 
seen as a source of turbulence for the conducting fluids described by incompressible magnetohydrodynamics (MHD), 
whereas the two others are more or less balanced (Els\"asser number of order one). This balance leads to the strong-field 
regime -- the so-called magnetostrophic dynamo -- for which we may derive magnetostrophic waves \citep{lehnert,schmitt08}. 
This regime is thought to be relevant not only for Earth but also for giant planets like Jupiter or Saturn, and by 
extension probably to exoplanets \citep{stevenson}. In order to investigate the dynamo problem several experiments 
have been developed \citep{ens}. 
In one of them, the authors were able to successfully reproduce with liquid sodium reversals and excursions of a turbulent 
dynamo generated by two (counter) rotating disks \citep{Berhanu}. This result follows a three-dimensional numerical simulation 
of the Earth's outer core where the reversal of the dipole moment was also obtained \citep{Glatzmaier}. In this model, however, 
the inertial/advection terms are simply discarded to mimic a very small Rossby number. This assumption is in apparent contradiction 
with any turbulent regime (Reynolds number is about $10^9$ for the Earth's outer core) and in particular with the weak 
turbulence one in which the nonlinear interactions -- although weak at short-time scales compared with the linear contributions 
-- become important for the nonlinear dynamics at asymptotically large-time scales. As we will see below, it is basically the 
regime that we shall investigate theoretically in this paper: a sea of helical (magnetized) waves \citep{moffatt70} will be 
considered as the main ingredient for the triggering of dynamo through the nonlinear transfer of magnetic energy and helicity. 

Weak turbulence is the study of the long time statistical behavior of a sea of weakly nonlinear dispersive waves 
\citep{nazarenko11}. The energy transfer between waves occurs mostly among resonant sets of waves and the resulting 
energy distribution, far from a thermodynamic equilibrium, is characterized by a wide power law spectrum and a high 
Reynolds number. This range of wavenumbers -- the inertial range -- is generally localized between large-scales at which 
energy is injected in the system (sources) and small-scales at which waves break or dissipate (sinks). Pioneering works on 
weak turbulence date back to the sixties when it was established that the stochastic initial value problem for weakly 
coupled wave systems has a natural asymptotic closure induced by the dispersive nature of the waves and the large 
separation of linear and nonlinear time scales \citep{benney66,BN67,benney69}. In the meantime, \cite{Zakh66} 
showed that the kinetic equations derived from the weak turbulence analysis have exact equilibrium solutions which are 
the thermodynamic zero flux solutions but also -- and more importantly -- finite flux solutions which describe the transfer of 
conserved quantities between sources and sinks. The solutions, first published for isotropic turbulence \citep{Zakh65,Zakh66} 
were then extended to anisotropic turbulence \citep{kuznetsov72}.  Weak turbulence 
is a very common natural regime with applications, for example, to capillary waves \citep{Kolmakov}, 
gravity waves \citep{Falcon}, superfluid helium and processes of Bose-Einstein condensation \citep{lvov03}, nonlinear 
optics \citep{Dyachenko}, inertial waves \citep{galtier03a}, Alfv\'en waves \citep{galtier00,galtier02,galtierchandran} or
whistler/kinetic Alfv\'en waves \citep{galtier06}. 

In this paper, the weak turbulence theory will be established for rotating MHD in the limit of small Rossby and Ekman 
numbers, the latter measuring the ratio of the viscous on Coriolis terms. We shall assume the existence of a strong 
uniform magnetic field parallel to the fast and constant rotating rate. The combination of the Coriolis and Lorentz-Laplace 
forces leads to the appearance of two types of circularly polarized waves and a possible non equipartition between the kinetic and 
magnetic energies \citep{moffatt72,favier}. After a general introduction to rotating MHD in \S \ref{sec2}, a weak helical turbulence 
formalism is developed in \S \ref{sect3} by using a technique developed in \cite{galtier06}. The phenomenology of weak 
turbulence dynamo is given in \S \ref{sect4bis}, the general properties of the weak turbulence equations are discussed 
in \S \ref{sect4}, whereas the exact spectral solutions are derived in \S \ref{sect5}. We conclude with a discussion in 
\S \ref{sect7}. Generally speaking, it is believed that the present work can be useful for better understanding the nonlinear 
magnetostrophic dynamo \citep{Roberts} with in background an application to Earth but also to giant planets like Jupiter 
or Saturn for which the intensity of the Coriolis force is relatively strong, the range of available length scales wide, and the 
magnetic field mainly dipolar with a weak tilt ($\le 10^{\rm o}$) of the dipole relative to the rotation axis. 
By extension we may even think that the analysis is relevant for exoplanets and some magnetized stars \citep{morin}.

\section{Rotating magnetohydrodynamics} \label{sec2}
\subsection{Governing equations}

The basic equations governing incompressible MHD under solid rotation and in the presence of a uniform
background magnetic field are:
\ba
\frac{\partial \uu}{\partial t} + 2 \Ome \times \uu + \uu \cdot \nabla \, \uu &=& 
- {\bf \nabla} P_* + \bbl \cdot \nabla \, \bb + \bb \cdot \nabla \, \bb + \nu \nabla^2 \uu \, , \label{rmhd1} \\
\frac{\partial \bb}{\partial t} + \uu \cdot \nabla \, \bb &=& \bbl \cdot \nabla \, \uu +
\bb \cdot \nabla \, \uu + \eta {\nabla}^2\bb \, , \label{rmhd2} \\
\nabla \cdot \uu &=& 0 \, , \label{rmhd3} \\
\nabla \cdot \bb &=& 0 \, , \label{rmhd4}
\ea
with $\uu$ the velocity, $P_*$ the total pressure (including the magnetic pressure and the centrifugal term), 
$\bb$ the magnetic field normalized to a velocity ($\bb \to \sqrt{\mu_0 \rho_0} \, \bb$, with $\rho_0$ the 
constant density), $\bbl$ the uniform normalized magnetic field, $\Omega_0$ the rotating rate, $\nu$ the 
kinematic viscosity and $\eta$ the magnetic diffusivity. Note the presence of the Coriolis force in the first 
equation (second term in the left hand side). Turbulence can only be maintained if a source is added to 
balance the small-scale dissipation. For example, in the geodynamo problem we may think that this external 
forcing is played by the convection (since the Rayleigh number $\sim 10^9$) with the buoyancy force \citep{braginsky}. 
In our case, we shall perform a pure nonlinear analysis, therefore the source and dissipation terms will be 
discarded. The weak turbulence equations that will be derived may describe, however, {\it any} magnetic 
Prandtl limit since the (linear) dissipative terms may be added to the equations after having made the 
nonlinear asymptotic analysis. 
In the rest of the paper, we shall assume that:
\be
\Ome = \Omes \ep \, , \quad \bbl = b_0 \ep \, ,
\ee
with $\ep$ a unit vector ($|\ep|=1$). We introduce the magneto-inertial length $d$ defined as:
\be
d \equiv {b_0 \over \Omes} \, . 
\ee
This length scale will be useful to characterize the main properties of rotating MHD.

\subsection{Three-dimensional inviscid invariants}
\label{invariants}

The two inviscid ($\nu=\eta=0$) quadratic invariants of incompressible rotating MHD in the presence of a 
background magnetic field parallel to the rotating axis are the total energy:
\be
E = \frac{1}{2} \int ( \uu^2 + \bb^2 ) \, d\,{\cal V} \, ,
\label{I1}
\ee
and the hybrid helicity:
\be
H = \frac{1}{2} \int \left( \uu \cdot \bb - {{\bf a} \cdot \bb \over d} \right) \, d\,{\cal V} \, ,
\label{I2}
\ee
where ${\bf a}$ is the vector potential ($\bb = \nabla \times {\bf a}$) and ${\cal V}$ is the volume over which 
the average is made. The second invariant is a mixture of cross-helicity, 
$H^c = (1/2) \int ( \uu \cdot \bb)\, d\,{\cal V}$, and magnetic helicity, 
$H^m = (1/2) \int ({\bf a} \cdot \bb)\, d\,{\cal V}$, which are not conserved in the present situation \citep{matthaeus82}. 
Indeed, it is straightforward to show from (\ref{rmhd1})--(\ref{rmhd4}) that \citep[see also][]{shebalin06}:
\ba
{\partial E \over \partial t} &=& - \int (\nu \, \ww^2 + \eta \, \jj^2) \, d\,{\cal V} \, , \\
{\partial H^c \over \partial t} &=& \Ome \cdot \int (\bb \times \uu) \, d\,{\cal V}
- (\nu+\eta) \int (\jj \cdot \ww) \, d\,{\cal V} \, , \\
{\partial H^m \over \partial t} &=& \bbl \cdot \int (\bb \times \uu) \, d\,{\cal V} 
- 2 \eta \int (\jj \cdot \bb) \, d\,{\cal V} \, , 
\ea
where $\ww$ is the vorticity and $\jj$ is the normalized current density. Therefore, the previous equations 
demonstrate that a second invariant may emerge {\it if and only if} $\bbl = d \Ome$. Below, we will verify that 
for the weak turbulence equations these two inviscid invariants are conserved for each triad of wave vectors.

\subsection{Helical MHD waves}

\begin{figure}
\centerline{\includegraphics[width=13cm]{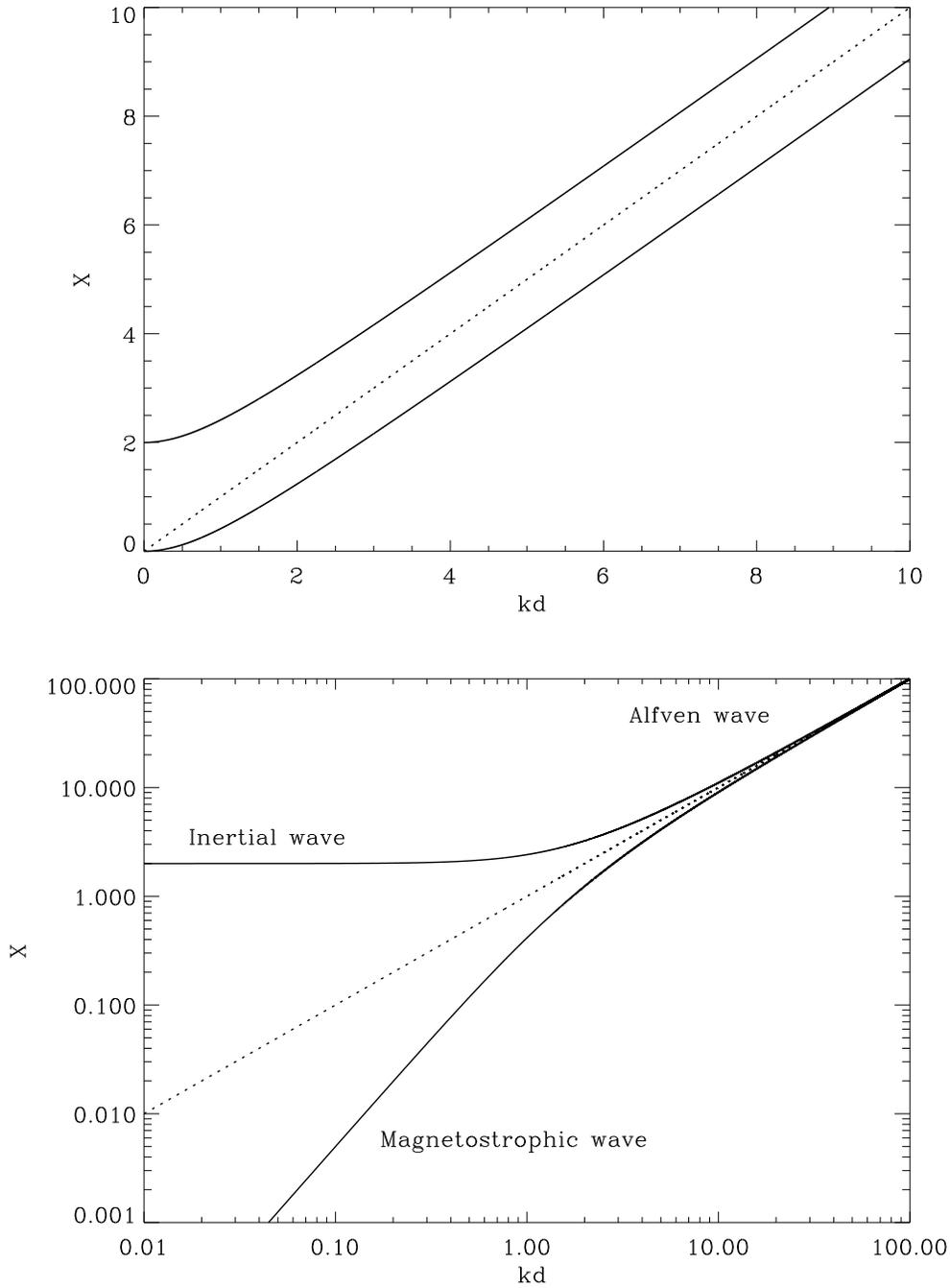}}
\caption{Dispersion relation for rotating MHD permeated by a background magnetic field in linear (top) and logarithmic 
(bottom) coordinates, with $X \equiv k \ols / (s \kpa \Omes)$. The upper and lower branches correspond respectively to 
left- and right-handed polarized waves. The Alfv\'en wave dispersion relation is also given (dotted line).}
\label{fig1}
\end{figure}

One of the main effects produced by the Coriolis force is to modify the polarization of the linearly polarized Alfv\'en 
waves -- solutions of the standard MHD equations -- which become circularly polarized and dispersive \citep{lehnert}. 
Indeed, if we linearize equations (\ref{rmhd1})--(\ref{rmhd4}) such that:
\be
\bb ({\bf x}) = \epsilon \, \bb ({\bf x}) \, , \quad 
\uu ({\bf x}) = \epsilon \, \uu ({\bf x}) \, , 
\ee
with $\epsilon$ a small parameter ($0<\epsilon \ll 1$) and ${\bf x}$ a three-dimensional displacement vector, 
then we obtain the following inviscid ($\nu=0$) and ideal ($\eta=0$) equations in Fourier space:
\ba
\partial_t \ww_\kk -2 i \kpa \Omes \uu_\kk - i \kpa b_0 \jj_\kk &=& 
\epsilon \, \{ \ww \cdot \nabla \, \uu - \uu \cdot \nabla \, \ww 
+ \bb \cdot \nabla \, \jj - \jj \cdot \nabla \, \bb \}_\kk \, , \label{wave1} \\
\partial_t \bb_\kk - i \kpa b_0 \uu_\kk &=& 
\epsilon \, \{ \bb \cdot \nabla \, \uu - \uu \cdot \nabla \, \bb \}_\kk \, , \label{wave2} \\
\kk \cdot \uu_\kk &=& 0 \, , \label{wave3} \\ 
\kk \cdot {\bf b}_\kk &=& 0 \, , \label{wave4}
\ea
where the wave vector $\kk = k\ek = \kkp + \kpa \ep$ ($k=|\kk|$, $\kpn = |\kkp|$, $|\ek|=1$) and $i^2 = -1$. 
The index $\kk$ denotes the Fourier transform, defined by the relation:
\be
\uu ({\bf x}) \equiv \int \uu (\kk) \, e^{i \kk \cdot {\bf x}} \, d\kk \, ,
\ee
where $\uu (\kk) = \uu_\kk = {\tilde \uu}_\kk e^{-i \omega t}$ (the same notation is used for the other fields). 
The linear dispersion relation ($\epsilon=0$) reads: 
\be
\omega^2 + \left({2 \Omes \, \kpa \over \Lambda k}\right) \, \omega - \kpa^2 b_0^2 = 0 \, ,
\label{dispersion}
\ee
with:
\be 
{{\tilde \uu}_\kk \brace {\tilde \bb}_\kk} = \Lambda \, i \, \ek \times {{\tilde \uu}_\kk \brace {\tilde \bb}_\kk} \, .
\label{pol1}
\ee
We obtain the general solution:
\be
\omega \equiv \ols = {s \kpa \Omes \over k} \left(- s \Lambda + \sqrt{1 + k^2 d^2} \right) \, , 
\label{dispersion2}
\ee
where the value ($\pm 1$) of $s$ defines the directional wave polarity such that we always have 
$s \kpa \ge 0$; then $\ols$ is a positive definite pulsation. The wave polarization $\Lambda$ tells us if the 
wave is right ($\Lambda=s$) or left ($\Lambda=-s$) circularly polarized. In the first case, we are dealing with 
the magnetostrophic branch, whereas in the latter case with the inertial branch (see figure \ref{fig1}). We see 
that the transverse circularly polarized waves are dispersive and that we recover the two well-known limits, 
\ie the pure inertial waves ($\omega_{-s}^s = 2 s \Omes \kpa / k \equiv \omega_I$) in the large-scale limit 
($kd \to 0$), and the standard Alfv\'en waves ($\omega = s \kpa b_0 \equiv \omega_A$) in the small-scale 
limit ($kd \to + \infty$). For the pure magnetostrophic waves we find the pulsation, 
$\omega_s^s = s \kpa k d b_0/2 = \omega^2_A / \omega_I \equiv \omega_M$. 
Note that the Alfv\'en waves become linearly polarized only when the Coriolis force vanishes: when it is 
present, whatever its magnitude is, the modified Alfv\'en waves are circularly polarized. This property is also 
found in MHD when the Hall term is added \citep{sahraoui07}. 

\subsection{Polarization}
\label{pol}

The polarizations $s$ and $\Lambda$ can be related to two well-known quantities, the reduced magnetic 
helicity $\sigma^m$ and the reduced cross-helicity $\sigma^c$. The reduced magnetic helicity is defined as:
\be
\sigma^m = { {\bf a}_\kk \cdot \bb_\kk^* + {\bf a}^*_\kk \cdot \bb_\kk \over 
2 \vert {\bf a}_\kk \vert \vert \bb_\kk \vert } \, ,
\ee
where $*$ denotes the complex conjugate. For circularly polarized waves, we can use relation (\ref{pol1}) 
which gives $\sigma^m = \Lambda$. On the other hand, the reduced cross-helicity is defined as:
\be
\sigma^c = { \uu_\kk \cdot \bb_\kk^* + \uu^*_\kk \cdot \bb_\kk \over 
2 \vert \uu_\kk \vert \vert \bb_\kk \vert } \, . 
\ee
The linear solution implies, $\omega \bb_\kk = - s \vert \kpa \vert b_0 \uu_\kk$, which leads to $\sigma^c = -s$. 
The use of both relations gives eventually:
\be
\sigma^m \sigma^c = - \Lambda s \, . 
\ee
This result is only valid for the linear solutions but may be generalized to any fluctuations in order to find the 
properties of helical turbulence \citep{Meyrand12}.

\subsection{Magnetostrophic equation}

The governing equations of rotating MHD can also be written in the following form:
\ba
\frac{\partial \ww}{\partial t} &=& \nabla \times \left[ \uu \times (\ww + 2 \Ome) + \jj \times (\bb + d \Ome) \right]
+ \nu \nabla^2 \ww \, , \label{magn1} \\
\frac{\partial \bb}{\partial t} &=& \nabla \times \left[ \uu \times (\bb + \bbl) \right] + \eta {\nabla}^2\bb \, , \label{magn2} 
\ea
where the relation $\bbl = d \Ome$ has been introduced. The magnetostrophic regime corresponds to a balance
between the Coriolis and the Lorentz-Laplace forces \citep{Finlay}. If we balance such terms in the linear case, we obtain 
the relation:
\be
2 \uu = - d \jj \, , 
\ee
which can be introduced in equation (\ref{magn2}) to give: 
\be 
\frac{\partial \bb}{\partial t} = - {d \over 2} \nabla \times \left[ (\nabla \times \bb) \times (\bb + \bbl) \right] 
+ \eta {\nabla}^2\bb \, . \label{magn3} 
\ee
Expression (\ref{magn3}) is the magnetostrophic equation which describes the nonlinear evolution of the 
magnetic field when both the rotation and the uniform magnetic field are relatively strong. It is asymptotically 
true in the sense that it only corresponds to the lower part of the magnetostrophic branch shown in figure 
\ref{fig1}. We may note 
immediately the similarity with the electron MHD equation introduced in plasma physics \citep{Kingsep} 
to describe the small space-time evolution of a magnetized plasma. The difference resides in the coefficient 
$d/2$ which is the ion skin depth $d_i$ in electron MHD. Then, it is not surprising that the linear solution gives 
the same (up to a factor $1/2$) dispersion relation as for whistler waves which are also right circularly polarized. 
We will see in section \ref{emhd} that indeed the general weak turbulence equations gives in the large-scale 
right-polarization limit the same equation (up to a factor) as in the electron MHD case \citep{galtier03b}.

\subsection{Complex helicity decomposition}
\label{chd}

Given the incompressibility constraints ($\ref{wave3}$) and ($\ref{wave4}$), it is convenient to project the 
rotating MHD equations in a plane orthogonal to $\kk$. We will use the complex helicity decomposition 
technique which has been shown to be effective in providing a compact description of the dynamics of 
three-dimensional incompressible fluids 
\citep{craya58,Kraichnan73,waleffe92,lesieur97,turner00,galtier03a,galtier06}.
The complex helicity basis is also particularly useful since it allows to diagonalize systems dealing with 
circularly polarized waves. We introduce the complex helicity decomposition:
\be
{\bf h^{\Lambda}}(\kk) \equiv \hh = \ete + i \Lambda \efi \, ,
\label{basis0}
\ee
where:
\be
\ete = \efi \times \ek \, , \quad \efi = \frac{\ep \times \ek}{|\ep \times \ek|} \, ,
\label{efi}
\ee
and $|\ete (\kk)|$=$|\efi(\kk)|$=$1$. We note that ($\ek$, $h^+_{\kk}$, $h^-_{\kk}$) form a complex basis with 
the following properties:
\ba
{\bf h^{-\Lambda}_{k}} &=& {\bf h^{\Lambda}_{-k}} \, , \\
\ek \times \hh &=& - i \Lambda \, \hh \, , \\
\kk \cdot \hh &=& 0 \, , \\
{\bf h^{\Lambda}_k} \cdot {\bf h^{\Lp}_k} &=& 2 \, \delta_{-\Lp \Lambda}\, .
\ea
We project the Fourier transform of the original vectors $\uu ({\bf x})$ and $\bb ({\bf x})$ on the helicity basis (see also 
Appendix \ref{hdecomp}):
\ba
\uu_\kk &=& \sum_{\Lambda} \, {\cal U}_{\Lambda} (\kk) \, {\bf h^{\Lambda}_k} 
= \sum_{\Lambda} \, {\cal U}_{\Lambda} \, {\bf h^{\Lambda}_k} \, , \\
\bb_\kk &=& \sum_{\Lambda} \, {\cal B}_{\Lambda} (\kk) \, {\bf h^{\Lambda}_k} 
= \sum_{\Lambda} \, {\cal B}_{\Lambda} \, {\bf h^{\Lambda}_k} \, ,
\label{basis1}
\ea
and in particular, we note that:
\ba
\ww_\kk &=& k \sum_{\Lambda} \Lambda \, {\cal U}_{\Lambda} \, {\bf h^{\Lambda}_k} \, , \\
\jj_\kk &=& k \sum_{\Lambda} \Lambda \, {\cal B}_{\Lambda} \, {\bf h^{\Lambda}_k} \, . 
\label{basis2}
\ea
We introduce the expressions of the new fields into the rotating MHD equations written in Fourier space and we 
multiply it by the vector ${\bf h^{\Lambda}_{-k}}$. First, we will focus on the linear dispersion relation 
($\epsilon=0$) which reads:
\be
\partial_t {\cal Z}_{\Lambda}^s = - i \, \ols {\cal Z}_{\Lambda}^s \, ,
\label{zdispersion}
\ee
with:
\ba
{\cal Z}_{\Lambda}^s &\equiv& {\cal U}_{\Lambda} + \xi_{\Lambda}^s {\cal B}_{\Lambda} \, , \label{zab} \\
\xi_{\Lambda}^s &\equiv& \frac{- s kd}{\left( - s \Lambda + \sqrt{1 + k^2 d^2} \right)} \, . 
\ea
Equation ($\ref{zdispersion}$) shows that ${\cal Z}_{\Lambda}^s$ are the canonical variables for our system. 
These eigenvectors combine the velocity and the magnetic field in a non trivial way by a factor 
$\xi_{\Lambda}^s$ (with $\ols =-b_0 \kpa / \xi_{\Lambda}^s$). In the small-scale limit ($k d \to + \infty$), 
we see that $\xi_{\Lambda}^s \to - s$: the Els\"asser variables used in standard MHD are then recovered. In the 
large-scale limit ($k d \to 0$), we have $\xi_{\Lambda}^s \to -s \, kd/2$ for $\Lambda = -s$ (inertial waves), 
or $\xi_{\Lambda}^s \to (-2 s / kd)^{-1}$ for $\Lambda = s$ (magnetostrophic wave). Therefore, 
${\cal Z}_{\Lambda}^s$ can be seen as a generalization of the Els\"asser variables to rotating MHD. In the 
rest of the paper, we shall use the relation:
\be
{\cal Z}_{\Lambda}^s = (\alf - \alfm) \, \aak \, e^{-i \ols t} \, , 
\label{ZversusA}
\ee
where $\aak$ is the wave amplitude in the interaction representation for which we have, in the linear 
approximation, $\partial_t \aak = 0$. In particular, that means that weak nonlinearities will modify only slowly 
in time the helical MHD wave amplitudes. The coefficient in front of the wave amplitude is introduced in 
advance to simplify the algebra that we are going to develop.

\section{Helical weak turbulence formalism} \label{sect3}
\subsection{Fundamental equations}

We decompose the inviscid nonlinear MHD equations (\ref{wave1})--(\ref{wave2}) on the complex helicity 
basis introduced in the previous section. Then, we project the equations on the vector ${\bf h^{\Lambda}_{-k}}$. 
After simplifications we obtain: 
\be
\partial_t {\cal U}_{\Lambda} - {2 i \Lambda \Omes \kpa \over k} {\cal U}_{\Lambda} 
- i b_0 \kpa {\cal B}_{\Lambda} = 
\ee
$$
\frac{i \epsilon}{2 \Lambda k} \int \sum_{\Lambda_p, \Lambda_q} (p \Lambda_p - q \Lambda_q) 
({\cal U}_{\Lambda_p} {\cal U}_{\Lambda_q} - {\cal B}_{\Lambda_p} {\cal B}_{\Lambda_q})
(\qq \cdot {\bf h^{\Lambda_p}_p}) ({\bf h^{\Lambda_q}_q} \cdot {\bf h^{-\Lambda}_k})
\, \delta_{pq,k} \, d{\bf p} \, d{\bf q} \, , 
$$
and:
\be
\partial_t {\cal B}_{\Lambda} - i b_0 \kpa {\cal U}_{\Lambda} = 
\ee
$$
\frac{i \epsilon}{2} \int \sum_{\Lambda_p, \Lambda_q}
({\cal U}_{\Lambda_q} {\cal B}_{\Lambda_p} - {\cal U}_{\Lambda_p} {\cal B}_{\Lambda_q})
(\qq \cdot {\bf h^{\Lambda_p}_p}) ({\bf h^{\Lambda_q}_q} \cdot {\bf h^{-\Lambda}_k})
\, \delta_{pq,k} \, d{\bf p} \, d{\bf q} \, , 
$$
where $\delta_{pq,k}=\delta({\bf p} + {\bf q} - \kk)$. The delta distributions come from the Fourier transforms 
of the nonlinear terms. We introduce the generalized Els\"asser variables in the following way:
\ba
{\cal U}_{\Lambda} &=& \sum_{s} { - \alfm {\cal Z}_{\Lambda}^s \over \alf - \alfm} \, , \\
{\cal B}_{\Lambda} &=& \sum_{s} { {\cal Z}_{\Lambda}^s \over \alf - \alfm} \, .
\ea
Then, we obtain in the interaction representation (variable $\aak$): 
\be
\partial_t \aak = 
\frac{i \epsilon}{2} \int \sum_{\Lambda_p, \Lambda_q \atop s_p, s_q}
L{{\Lambda \Lambda_p \Lambda_q \atop s \, s_p \, s_q} \atop -k \, p \, q}
\, \aap \aaq \, e^{-i \Omega_{pq,k} t} \, \delta_{pq,k} \, d{\bf p} \, d{\bf q} \, ,
\label{fonda1}
\ee
where:
\be
L{{\Lambda \Lambda_p \Lambda_q \atop s \, s_p \, s_q} \atop k \, \, p \, \, q} = 
\label{Lgeo}
\ee
$$
\left[ \left({p \Lambda_p - q \Lambda_q \over \Lambda k} \right) \left(\alfmp \alfmq -1\right) + 
\alf \left(\alfmp - \alfmq\right) \right]
{\left(\qq \cdot {\bf h^{\Lambda_p}_p}\right) \left({\bf h^{\Lambda_q}_q} \cdot {\bf h^{\Lambda}_k}\right) 
\over \alf - \alfm} \, ,
$$
and:
\be
\Omega_{pq,k} = \olsp + \olsq - \ols \, .
\ee
Equation (\ref{fonda1}) is the wave amplitude equation from which it is possible to extract some information. 
As expected we see that the nonlinear terms are of order $\epsilon$. This means that weak nonlinearities 
will modify only slowly in time the helical MHD wave amplitude. They contain an exponentially oscillating 
term which is essential for the asymptotic closure. Indeed, weak turbulence deals with variations of spectral 
densities at very large time, \ie for a nonlinear transfer time much greater than the wave period. As a 
consequence, most of the nonlinear terms are destroyed by phase mixing and only a few of them, the 
resonance terms, survive (see \eg \cite{newell01}). The expression obtained for the fundamental 
equation (\ref{fonda1}) is classical in weak turbulence. The main difference between different problems is 
localized in the matrix $L$ which is interpreted as a complex geometric coefficient. We will see below that 
the local decomposition allows to get a polar form for such a coefficient which is much easier to manipulate. 
From equation (\ref{fonda1}) we see eventually that, contrary to incompressible MHD, there is no exact 
solutions to the nonlinear problem in incompressible rotating MHD. The origin of such a difference is that in 
MHD the nonlinear term involves Alfv\'en waves traveling only in opposite directions whereas in rotating 
MHD this constrain does not exist (we have a summation over $\Lambda$ and $s$). In other words, if one 
type of wave is not present in incompressible MHD then the nonlinear term cancels whereas in the present 
problem it is not the case \citep[see \eg][]{galtier00}.

\subsection{Local decomposition}
\label{local}

\begin{figure}
\centerline{\includegraphics[width=13cm]{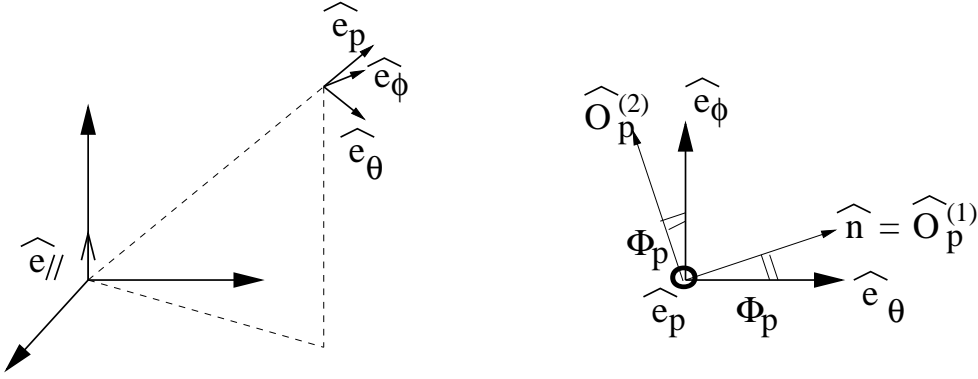}}
\caption{Local decomposition for a given wave vector $\pp$.}
\label{fig2}
\end{figure}

In order to evaluate the scalar products of complex helical vectors found in the geometric coefficient 
(\ref{Lgeo}), it is convenient to introduce a vector basis local to each particular triad 
\citep{waleffe92,turner00,galtier03a}. For example, for a given vector $\pp$, we define the orthonormal 
basis vectors:
\ba
{\bf {\hat O^{(1)}}}(\pp) &=& \nn \, ,\\
{\bf {\hat O^{(2)}}}(\pp) &=& \eep \times \nn \, ,\\
{\bf {\hat O^{(3)}}}(\pp) &=& \eep \, ,
\label{basis2}
\ea
where $\eep=\pp/|\pp|$ and:
\be
\nn = {\pp \times \kk \over | \pp \times \kk |} = {\qq \times \pp \over | \qq \times \pp |} =
{\kk \times \qq \over | \kk \times \qq |} \, .
\ee
We see that the vector $\nn$ is normal to any vector of the triad ($\kk$,$\pp$,$\qq$) and changes sign if 
$\pp$ and $\qq$ are interchanged, \ie $\nn_{(\kk,\qq,\pp)} = - \nn_{(\kk,\pp,\qq)}$. Note that $\nn$ does 
not change by cyclic permutation, \ie, $\nn_{(\kk,\qq,\pp)} = \nn_{(\qq,\pp,\kk)} =\nn_{(\pp,\kk,\qq)}$. 
A sketch of the local decomposition is given in figure \ref{fig2}. 
We now introduce the vectors:
\be
\Xi^{\Lambda_p}(\pp) \equiv \Xi^{\Lambda_p}_\pp = 
{\bf {\hat O^{(1)}}}(\pp) + i \Lambda_p {\bf {\hat O^{(2)}}}(\pp) \, ,
\ee
and define the rotation angle $\Phi_p$, so that:
\ba
\cos \Phi_p &=& \nn \cdot \ete (\pp) \, , \\
\sin \Phi_p &=& \nn \cdot \efi (\pp) \, .
\ea
The decomposition of the helicity vector ${\bf h^{\Lambda_p}_p}$ in the local basis gives (similar forms
are obtained for $\kk$ and $\qq$):
\be
{\bf h^{\Lambda_p}_p} = \Xi^{\Lambda_p}_\pp \, e^{i \Lambda_p \Phi_p} \, . 
\ee
After some algebra we obtain the following polar form for the matrix $L$:
\be
L{{\Lambda \Lambda_p \Lambda_q \atop s \, s_p \, s_q} \atop k \, \, p \, \, q} = 
-\left[ \left({p \Lambda_p - q \Lambda_q \over \Lambda k} \right) 
\left(\alfmp \alfmq -1\right) + \alf \left(\alfmp - \alfmq\right) \right]
\label{lint2}
\ee
$$
i e^{i (\Lambda \Phi_k + \Lambda_p \Phi_p + \Lambda_q \Phi_q)} \, 
\frac{\Lambda \, \Lambda_p \, \Lambda_q}{\alf - \alfm} \, \frac{\sin \psi_k}{k} \, 
k q \, (\Lambda \Lambda_q + \cos \psi_p) \, .
$$
The angle $\psi_k$ refers to the angle opposite to $\kk$ in the triangle defined by $\kk=\pp+\qq$ 
($\sin \psi_k = \nn\cdot(\qq \times \pp)/|(\qq \times \pp)|$). To obtain equation (\ref{lint2}), we have also 
used the well-known triangle relations:
\be
{\sin \psi_k \over k} = {\sin \psi_p \over p} = {\sin \psi_q \over q} \, . 
\ee
Further modifications have to be made before applying the spectral formalism. In particular, the fundamental 
equation has to be invariant under interchange of $\pp$ and $\qq$. To do so, we shall introduce the 
symmetrized matrix:
\be
{1 \over 2} \left( L{{\Lambda \Lambda_p \Lambda_q \atop s \, s_p \, s_q} \atop k \, \, p \, \, q}  + 
L{{\Lambda \Lambda_q \Lambda_p \atop s \, s_q \, s_p} \atop k \, \, q \, \, p} \right) \, .
\ee
Finally, by using the identities given in Appendix \ref{relation}, we obtain:
\be
\partial_t \aak = 
\frac{\epsilon \, d^2}{16} \int \sum_{\Lambda_p, \Lambda_q \atop s_p, s_q}
\frac{\alfmq - \alfmp}{\alf - \alfm} \, 
M{{\Lambda \Lambda_p \Lambda_q \atop s \, s_p \, s_q} \atop -k \, p \, q}
\, \aap \aaq \, e^{-i \Omega_{pq,k} t} \, \delta_{pq,k} \, d{\bf p} \, d{\bf q} \, ,
\label{fonda2}
\ee
where:
\be
M{{\Lambda \Lambda_p \Lambda_q \atop s \, s_p \, s_q} \atop k \, \, p \, \, q} = 
e^{i (\Lambda \Phi_k + \Lambda_p \Phi_p + \Lambda_q \Phi_q)} \, 
(\Lambda k + \Lambda_p p + \Lambda_q q) kpq \frac{\sin \psi_k}{k}
\label{fonda2bis}
\ee
$$
\alf \alfp \alfq \left(2 + {\alfm}^2 {\alfmp}^2 {\alfmq}^2 - {\alfm}^2 - {\alfmp}^2 - {\alfmq}^2 \right) \, .
$$
The matrix $M$ possesses the following properties:
\be
\left(
M{{\Lambda \Lambda_p \Lambda_q \atop s \, s_p \, s_q} \atop k \, \, p \, \, q}
\right)^* = 
M{{-\Lambda -\Lambda_p -\Lambda_q \atop -s \, -s_p \, -s_q} \atop k \, \, \, p \, \, \, q} =
M{{\Lambda \, \, \, \, \Lambda_p \, \, \, \, \Lambda_q \atop s \, \, \, \, s_p 
\, \, \, \, s_q} \atop -k-p-q} \, , 
\label{prop1}
\ee
\be
M{{\Lambda \Lambda_p \Lambda_q \atop s \, s_p \, s_q} \atop k \, \, p \, \, q} = 
- M{{\Lambda \Lambda_q \Lambda_p \atop s \, s_q \, s_p} \atop k \, \, q \, \, p} \, ,
\label{prop2}
\ee
\be
M{{\Lambda \Lambda_p \Lambda_q \atop s \, s_p \, s_q} \atop k \, \, p \, \, q} = 
- M{{\Lambda_q \Lambda_p \Lambda \atop s_q \, s_p \, s} \atop q \, \, p \, \, k} \, ,
\label{prop3}
\ee
\be
M{{\Lambda \Lambda_p \Lambda_q \atop s \, s_p \, s_q} \atop k \, \, p \, \, q} = 
-M{{\Lambda_p \Lambda \Lambda_q \atop s_p \, s \, s_q} \atop p \, \, k \, \, q} \, .
\label{prop4}
\ee
Equation (\ref{fonda2}) is the fundamental equation that describes the slow evolution of the wave amplitudes 
due to the nonlinear terms of the incompressible rotating MHD equations. It is the starting point for deriving 
the weak turbulence equations. The local decomposition used here allows us to represent concisely complex 
information in an exponential function (polar form). As we will see below, it will simplify significantly the derivation 
of the asymptotic equations. 

From equation (\ref{fonda2}) we note that the nonlinear coupling between helicity states associated with 
wave vectors, $\pp$ and $\qq$, vanishes when the wave vectors are collinear (since then, $\sin \psi_k=0$). 
This property is similar to the one found for pure rotating hydrodynamics. It seems to be a general property 
for helical waves \citep{Kraichnan73,waleffe92,turner00,galtier03a,galtier06}.
Additionally, we note that the nonlinear coupling between helicity states vanishes whenever the wave 
numbers $p$ and $q$ are equal if their associated wave and directional polarities, $\Lambda_p$, 
$\Lambda_q$, and $s_p$, $s_q$ respectively, are also equal. In the case of inertial waves, for which we 
have $\Lambda=-s$ (left-handed waves), this property was already observed \citep{galtier03a}. Here, this 
finding is generalized to right and left circularly polarized waves. Note that in the large-scale limit for which 
we recover the linearly polarized Alfv\'en waves, this property tends to disappear (see also section \ref{sec44}).

We are interested by the long-time behavior of the helical wave amplitudes. From the fundamental equation 
(\ref{fonda2}), we see that the nonlinear wave coupling will come from resonant terms such that:
\be
\left\{
\begin{array}{lll}
\kk = \pp + \qq \, , \\[.2cm]
{\kpa \over \alf} = {p_{\pa} \over \alfp} + {q_{\pa} \over \alfq} \, . 
\end{array}
\right.
\label{resonance0}
\ee
The resonance condition may also be written:
\be
\frac{\alfm - \alfmp}{q_{\pa}} = 
\frac{\alfmq - \alfm}{p_{\pa}} =
\frac{\alfmq - \alfmp}{k_{\pa}} \, .
\label{resonance}
\ee
As we shall see below, relations (\ref{resonance}) are useful in simplifying the weak turbulence equations and 
demonstrating the conservation of inviscid invariants.

\subsection{Asymptotic weak turbulence equations}

Weak turbulence is a state of a system composed of many simultaneously excited and interacting nonlinear 
waves where the energy distribution, far from thermodynamic equilibrium, is characterized by a wide power 
law spectrum. This range of wave numbers -- the inertial range -- is generally localized between large-scales 
at which energy is injected in the system and small dissipative scales. The origin of weak turbulence dates 
back to the early sixties and since then many papers have been devoted to the subject (see \eg 
\cite{hasselman62,benney66,zakharov67,sagdeev69,kuznetsov72,ZLF92,Galtier09,nazarenko11}).
The essence of weak turbulence is the statistical study of large ensembles of weakly interacting dispersive 
waves {\it via} a systematic asymptotic expansion in powers of small nonlinearity. This technique leads finally 
to the derivation of kinetic equations for quantities like the energy and more generally for the (quadratic) 
invariants of the system under investigation. Here, we will follow the standard Eulerian formalism of weak 
turbulence (see \eg \cite{benney69}).

We define the density tensor $\qls(\kk)$ for an homogeneous turbulence, such that:
\be
\langle \aak(\kk) \, \aakp(\kkpr) \rangle \equiv 
\qls(\kk) \, \delta (\kk + \kkpr) \, \delta_{\Lambda \Lambda^{\prime}} \, 
\delta_{s s^{\prime}} \, , 
\label{tensor1}
\ee
for which we shall write an asymptotic closure equation. The presence of the deltas 
$\delta_{\Lambda \Lambda^{\prime}}$ and $\delta_{s s^{\prime}}$ means that correlations with opposite wave 
or directional polarities have no long-time influence in the wave turbulence regime; the third delta distribution 
$\delta (\kk + \kkpr)$ is the consequence of the homogeneity assumption. Details of the derivation of the weak 
turbulence equations are given in Appendix \ref{derivation}. After a lengthly calculation, we obtain the following 
result: 
\be
\partial_t \qls (\kk) = 
\label{ke1}
\ee
$$
\frac{\pi \, \epsilon^2 d^4}{64 \, b_0^2} \int \sum_{\Lambda_p, \Lambda_q \atop s_p, s_q} 
\left( \frac{\sin \psi_k}{k} \right)^2 k^2 p^2 q^2 (\Lambda k + \Lambda_p p + \Lambda_q q)^2 
{\alf}^2 {\alfp}^2 {\alfq}^2 \left( \frac{\alfmq - \alfmp}{\kpa} \right)^2
$$
$$
\left( 2 + {\alfm}^2 {\alfmp}^2 {\alfmq}^2 - {\alfm}^2 - {\alfmp}^2 - {\alfmq}^2 \right)^2 
\left( \frac{\ols}{1+{\alfm}^2} \right) \qls (\kk) \, \qlsp (\pp) \, \qlsq (\qq)
$$
$$
\left[ {\ols \over (1+ {\alfm}^2) \qls(\kk)} - {\olsp \over (1+{\alfmp}^2) \qlsp(\pp)} - 
{\olsq \over (1+{\alfmq}^2) \qlsq(\qq)} \right] 
\delta(\Omega_{k,pq})\, \delta_{k,pq} \, d{\bf p} \, d{\bf q} \, . 
$$
Equation (\ref{ke1}) is the main result of the helical weak turbulence formalism. It describes the statistical 
properties of weak turbulence for rotating MHD at the lowest order, \ie for three-wave interactions.

\section{Phenomenology of weak turbulence dynamo} \label{sect4bis}

\begin{figure}
\centerline{\includegraphics[width=13cm]{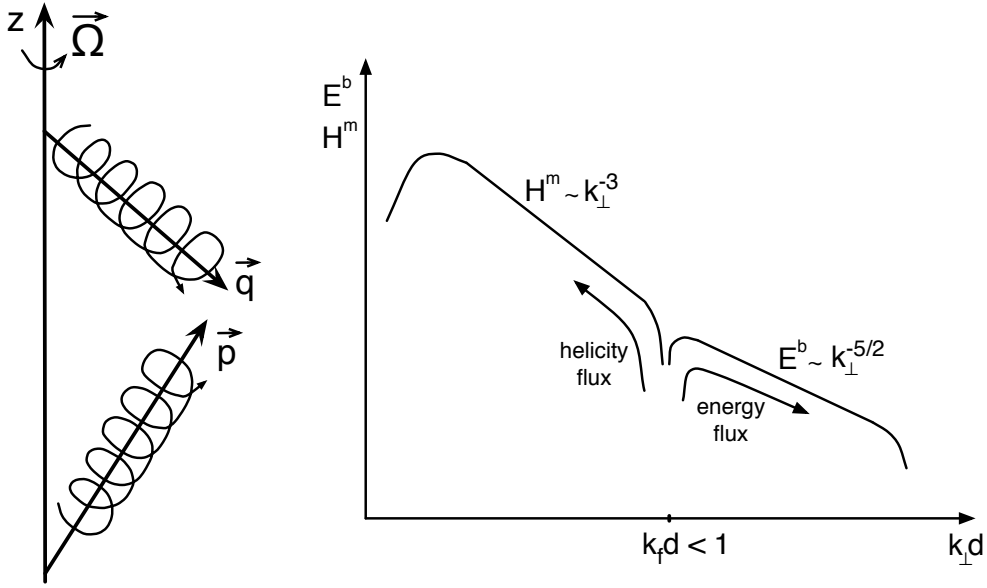}}
\caption{Left: collision between counter propagating circularly polarized waves.
Right: heuristic view of the magnetic energy and hybrid helicity spectra with a forcing applied at an intermediate scale $k_fd<1$; 
the inverse cascade of helicity may also drive the energy to the largest scales of the system.}
\label{fig3}
\end{figure}

Before going to the detailed analysis of the weak turbulence regime, it is important to have a simple picture in mind of 
the physical process that we are going to describe. According to the properties given in Section \ref{local}, if we assume 
that the nonlinear transfer is mainly driven by local interactions ($k \sim p \sim q$), then we may only consider the stochastic 
collisions between counter propagating waves \citep{iroshnikov64,kraichnan65} 
of the same kind to derive the form of the energy spectra (see figure \ref{fig3}): 
in other words, a left (right) handed wave going upward will interact much stronger with another left (right) handed wave 
propagating downward than one going upward. 

To find the transfer time and then the energy spectrum, we first need to evaluate the modification of a wave produced 
by one collision. Starting from the momentum equation (for simplicity we note the wave amplitude ${\cal Z}_\ell$ and 
we assume anisotropy with $k \sim \kpn$):
\be
{\cal Z}_\ell(t+\tau_1) \sim {\cal Z}_\ell(t) + \tau_1 {\partial {\cal Z}_\ell \over \partial t} 
\sim {\cal Z}_\ell(t) + \tau_1 {{\cal Z}^2_\ell \over \ell_\perp} \, ,
\ee
where $\tau_1$ is the duration of one collision; in other words, after one collision the distortion of a wave is 
$\Delta_1 {\cal Z}_\ell \sim \tau_1 {\cal Z}^2_\ell / \ell_\perp$. This distortion is going to increase with time in such a 
way that after $N$ stochastic collisions the cumulative effect may be evaluated like a random walk:
\be
\sum_{i=1}^N \Delta_i {\cal Z}_\ell \sim \tau_1 {{\cal Z}^2_\ell \over \ell_\perp} \sqrt{t \over \tau_1} \, .
\ee
The transfer time $\tau_{tr}$ that we are looking for is the one for which the cumulative distortion is of the order 
of one, \ie of the order of the wave itself:
\be
{\cal Z}_\ell \sim \tau_1 {{\cal Z}^2_\ell \over \ell_\perp} \sqrt{\tau_{tr} \over \tau_1} \, ,
\ee
then we obtain:
\be
\tau_{tr} \sim {1 \over \tau_1} {\ell_\perp^2 \over {\cal Z}^2_\ell} \sim {\tau^2_{NL} \over \tau_1} \, . 
\ee
It is basically the formula that we are going to use to evaluate the energy spectra. Let us consider inertial waves for which
$\tau_1 \sim 1/\omega_I$. A classical calculation, with $\varepsilon^u \sim {\cal Z}^2_\ell / \tau_{tr}$, leads finally to the 
bi-dimensional axisymmetric kinetic energy spectrum:
\be
E^u(\kpn,\kpa) \sim \sqrt{\varepsilon^u \Omega_0} \, \kpn^{-5/2} \kpa^{-1/2} \, ,
\ee
which is the prediction for weak inertial wave turbulence \citep{galtier03a}. Note that this solution corresponds to a 
constant kinetic energy flux $\varepsilon^u$ whereas a constant kinetic helicity flux may give other solutions 
\citep{galtier2014}. 
For magnetostrophic waves we have $\tau_1 \sim 1/\omega_M$, but a subtlety arrives because instead of the momentum 
equation now we use Eq. (\ref{magn3}) for which the nonlinear term leads to $\tau_{NL} \sim \ell_\perp^2 / (d {\cal Z}_\ell)$. 
Then, we obtain the bi-dimensional axisymmetric magnetic energy spectrum:
\be
E^b(\kpn,\kpa) \sim \sqrt{\varepsilon^b b_0 \over d} \, \kpn^{-5/2} \kpa^{-1/2} \, ,
\ee
which corresponds to a constant magnetic energy flux $\varepsilon^b$ solution. 

The same heuristic analysis can be made for the other invariant, the hybrid helicity. Let us consider the most interesting case, 
namely the magnetostrophic regime in which the hybrid helicity is mainly dominated by the magnetic helicity (with $kd<1$). 
By using the transfer time derived above (with the helicity flux ${\tilde \varepsilon} \sim H_\ell / \tau_{tr}$), we find:
\be
H(\kpn,\kpa) \sim H^m(\kpn,\kpa) \sim \sqrt{{\tilde \varepsilon} b_0 \over d^2} \, \kpn^{-3} \kpa^{-1/2} \, . 
\ee
An inverse cascade may happen for the hybrid helicity (see figure \ref{fig3}) which in turn may drive the magnetic energy 
at the largest scales of the system. 
It is through this mechanism that the large-scale magnetic field can be regenerated by the weak turbulence dynamo. It is fundamental 
to have in mind that this cascade happens because the hybrid helicity is an inviscid and ideal invariant of rotating MHD
(\eg without rotation an inverse cascade of magnetic helicity is impossible in weak incompressible MHD turbulence
\citep{galtierNaza}).
In other words, the inverse cascade should stop as soon as the mean magnetic field and the rotating rate are not collinear 
anymore. It is likely, however, that the inverse cascade is only weakly reduced when the mean magnetic field and the 
rotating rate experience a slight out of alignment (weak tilt case) and is completely inhibited in the strong tilt case.
This comment might explain why the planetary magnetic fields are often dipolar with a weak tilt ($< 10^{\rm o}$) of the 
dipole relative to the rotation axis. 
Note that the increase of the magnetic field at large-scale may lead to a state where the ratio between the magnetic and 
kinetic energies is significantly larger than one.

\section{General properties} \label{sect4}
\subsection{Basic turbulent spectra}

In section \ref{invariants}, we have introduced the three-dimensional inviscid invariants of incompressible 
rotating MHD. The first test that the weak turbulence equations have to satisfy is the detailed conservation 
of these invariants, that is to say the conservation of invariants for each triad ($\kk$, $\pp$, $\qq$). Starting 
from definitions (\ref{I1})--(\ref{I2}), we find the total energy spectrum: 
\be
E(\kk) = \sum_{\Lambda, s} (1 + {\alfm}^2) \, \qls (\kk) 
\equiv \sum_{\Lambda, s} {\cal E}_{\Lambda}^s (\kk) \, ,
\ee
which is composed of the magnetic spectrum:
\be
E^b(\kk) = \sum_{\Lambda, s} \qls (\kk) \, ,
\ee
and the kinetic spectrum:
\be
E^u(\kk) = \sum_{\Lambda, s} {\alfm}^2 \, \qls (\kk)  \, .
\ee
We also find the cross-helicity spectrum:
\be
H^c(\kk) = - \sum_{\Lambda, s} \alfm \, \qls (\kk) \, ,
\ee
and the magnetic helicity spectrum:
\be
H^m(\kk) = \sum_{\Lambda, s} \frac{\Lambda}{k} \, \qls (\kk) \, .
\ee
Note that each of these spectra may be decomposed into right ($\Lambda=s$) and left ($\Lambda=-s$) polarization spectra. 
From the last two expressions we find the second inviscid invariant, the hybrid helicity spectrum:
\be
H(\kk) = \sum_{\Lambda, s} \left( \frac{\alf - \alfm}{2} \right) \, \qls (\kk) 
\equiv \sum_{\Lambda, s} {\cal H}_{\Lambda}^s (\kk) \, .
\ee
We shall demonstrate below the conservation of the energy and the hybrid helicity.

\subsection{Triadic conservation of inviscid invariants}

We will first check the energy conservation. From expression (\ref{ke1}), we may write:
\be
\partial_t E (t) \equiv \partial_t \int E (\kk) \, d\kk \equiv 
\partial_t  \int \sum_{\Lambda, s} {\cal E}_{\Lambda}^s (\kk) \, d\kk = 
\label{kenergy0}
\ee
$$
\frac{\pi \, \epsilon^2 \, d^4}{64 \, b_0^2} \int 
\sum_{\Lambda, \Lambda_p, \Lambda_q \atop s, s_p, s_q} 
\left( \frac{\sin \psi_k}{k} \right)^2 k^2 p^2 q^2 (\Lambda k + \Lambda_p p + \Lambda_q q)^2 
{\alf}^2 {\alfp}^2 {\alfq}^2 \left( \frac{\alfmq - \alfmp}{\kpa} \right)^2
$$
$$
\left( 2 + {\alfm}^2 {\alfmp}^2 {\alfmq}^2 - {\alfm}^2 - {\alfmp}^2 - {\alfmq}^2 \right)^2 
\qls (\kk) \, \qlsp (\pp) \, \qlsq (\qq)
$$
$$
\ols \, \left[ \frac{\ols}{{\cal E}_{\Lambda}^s(\kk)} +
\frac{\olsp}{{\cal E}_{\Lambda_p}^{s_p}(\pp)} +
\frac{\olsq}{{\cal E}_{\Lambda_q}^{s_q}(\qq)} \right]
\delta(\Omega_{kpq})\, \delta_{kpq} \, d\kk \, d\pp \, d\qq \, . 
$$
Equation (\ref{kenergy0}) is invariant under cyclic permutations of wave vectors; it leads to: 
\be
\partial_t E (t) = 
\label{kenergy1}
\ee
$$
\frac{\pi \, \epsilon^2 \, d^4}{192 \, b_0^2} \int 
\sum_{\Lambda, \Lambda_p, \Lambda_q \atop s, s_p, s_q} 
\left( \frac{\sin \psi_k}{k} \right)^2 k^2 p^2 q^2 (\Lambda k + \Lambda_p p + \Lambda_q q)^2 
{\alf}^2 {\alfp}^2 {\alfq}^2 \left( \frac{\alfmq - \alfmp}{\kpa} \right)^2
$$
$$
\left( 2 + {\alfm}^2 {\alfmp}^2 {\alfmq}^2 - {\alfm}^2 - {\alfmp}^2 - {\alfmq}^2 \right)^2 
\qls (\kk) \, \qlsp (\pp) \, \qlsq (\qq)
$$
$$
\Omega_{kpq} \, \left[ \frac{\ols}{{\cal E}_{\Lambda}^s(\kk)} +
\frac{\olsp}{{\cal E}_{\Lambda_p}^{s_p}(\pp)} + \frac{\olsq}{{\cal E}_{\Lambda_q}^{s_q}(\qq)} \right]
\delta(\Omega_{kpq})\, \delta_{kpq} \, d\kk \, d\pp \, d\qq \, . 
$$
On the resonant manifold $\Omega_{kpq}=0$, therefore the total energy is conserved exactly for each triad:
we have a detailed conservation of the total energy. 

For the second invariant it is straightforward to show with relation (\ref{ident5}) that: 
\be
\partial_t H (t) \equiv \partial_t \int H (\kk) \, d\kk 
\equiv \partial_t \int \sum_{\Lambda, s} {\cal H}_{\Lambda}^s (\kk) \, d\kk = 
\label{khelicity1}
\ee
$$
\frac{\pi \, \epsilon^2 \, d^4}{64 \, b_0^2} \int 
\sum_{\Lambda, \Lambda_p, \Lambda_q \atop s, s_p, s_q} 
\left( \frac{\sin \psi_k}{k} \right)^2 k^2 p^2 q^2 (\Lambda k + \Lambda_p p + \Lambda_q q)^2 
{\alf}^2 {\alfp}^2 {\alfq}^2 \left( \frac{\alfmq - \alfmp}{\kpa} \right)^2
$$
$$
\left( 2 + {\alfm}^2 {\alfmp}^2 {\alfmq}^2 - {\alfm}^2 - {\alfmp}^2 - {\alfmq}^2 \right)^2 
\qls (\kk) \, \qlsp (\pp) \, \qlsq (\qq)
$$
$$
{\alf \over 2} \ols \, \left[ \frac{2 \alf \ols}{{\cal H}_{\Lambda}^s(\kk)} +
\frac{2 \alfp \olsp}{{\cal H}_{\Lambda_p}^{s_p}(\pp)} +
\frac{2 \alfq \olsq}{{\cal H}_{\Lambda_q}^{s_q}(\qq)} \right]
\delta(\Omega_{kpq})\, \delta_{kpq} \, d\kk \, d\pp \, d\qq \, . 
$$
Equation (\ref{kenergy0}) is also invariant under cyclic permutations of wave vectors. Then, one is led to:
\be
\partial_t H(t) = 
\label{khelicity2}
\ee
$$
\frac{\pi \, \epsilon^2 \, d^4}{192} \int 
\sum_{\Lambda, \Lambda_p, \Lambda_q \atop s, s_p, s_q} 
\left( \frac{\sin \psi_k}{k} \right)^2 k^2 p^2 q^2 (\Lambda k + \Lambda_p p + \Lambda_q q)^2 
{\alf}^2 {\alfp}^2 {\alfq}^2 \left( \frac{\alfmq - \alfmp}{\kpa} \right)^2
$$
$$
\left( 2 + {\alfm}^2 {\alfmp}^2 {\alfmq}^2 - {\alfm}^2 - {\alfmp}^2 - {\alfmq}^2 \right)^2 
\qls (\kk) \, \qlsp (\pp) \, \qlsq (\qq)
$$
$$
(\kpa + p_{\pa} + q_{\pa}) \, 
\left[ \frac{\kpa}{{\cal H}_{\Lambda}^s(\kk)} +
\frac{p_\pa}{{\cal H}_{\Lambda_p}^{s_p}(\pp)} +
\frac{q_\pa}{{\cal H}_{\Lambda_q}^{s_q}(\qq)} \right]
\delta(\Omega_{kpq})\, \delta_{kpq} \, d\kk \, d\pp \, d\qq \, ,
$$
which is exactly equal to zero on the resonant manifold: we also have the triadic conservation for the hybrid 
helicity.

\subsection{Helical properties}

From the weak turbulence equations (\ref{ke1}), we find several general properties. Some of them can be 
obtained directly from the wave amplitude equation (\ref{fonda2}) as explained in section \ref{local}. First, 
we observe that there is no coupling between helical waves associated with wave vectors, $\pp$ and $\qq$, 
when the wave vectors are collinear ($\sin \psi_k=0$). Second, we note that there is no coupling between 
helical waves associate with vectors $\pp$ and $\qq$ whenever their magnitudes, $p$ and $q$, are equal if 
their associated polarities, $s_p$ and $s_q$ in one hand and, $\Lambda_p$ and $\Lambda_q$ on the other 
hand, are also equal (since then $\alfmq - \alfmp=0$). These properties hold for the inviscid invariants and 
generalize what was found previously for rotating hydrodynamics \citep{galtier03a} where we only have 
left circularly polarized waves ($\Lambda=-s$). It seems to be a generic property of helical wave interactions 
\citep{Kraichnan73,waleffe92,turner00}. As noted before, this property tends to 
disappear when the large-scale limit is taken, \ie when we tend to the standard MHD. Third, it follows from 
the previous observations that a strong helical perturbation localized initially in a narrow band of wave 
numbers will lead to a weak transfer of total energy and hybrid helicities. Note that these properties can be 
inferred from the fundamental equation (\ref{fonda2}) as well.

\subsection{Small-scale dynamics: Alfv\'en waves}
\label{sec44}

We start with the general weak turbulence equation (\ref{ke1}) and take the small-scale limit ($kd \to + \infty$) 
for which we have, at the leading order:
\ba
\alf &\to& -s \, , \\
\left( \alfmq - \alfmp \right)^2 &\to& (s_q - s_p)^2 \, , \label{lim2} \\
\left( 2 + {\alfm}^2 {\alfmp}^2 {\alfmq}^2 - {\alfm}^2 - {\alfmp}^2 - {\alfmq}^2 \right)^2 &\to& 
{16 \over d^4} \left( {s \Lambda k + s_p \Lambda_p p + s_q \Lambda_q q \over kpq} \right)^2 , \quad \quad \\
\ols &\to& s \kpa b_0 = \omega_A \, .
\ea
After introducing the previous expressions into (\ref{ke1}), we obtain:
\be
\partial_t \qls (\kk) = 
\ee
$$
\frac{\pi \, \epsilon^2}{16 \, b_0} \int \sum_{\Lambda_p, \Lambda_q \atop s_p, s_q} 
\left( \frac{\sin \psi_k}{k} \right)^2 (\Lambda k + \Lambda_p p + \Lambda_q q)^2 
\left( \frac{s_q - s_p}{\kpa} \right)^2 \left( s \Lambda k + s_p \Lambda_p p + s_p \Lambda_q q \right)^2 s \kpa
$$
$$
\qls (\kk) \, \qlsp (\pp) \, \qlsq (\qq)
\left[ {s \kpa \over \qls(\kk)} - {s_p p_\pa \over \qlsp(\pp)} - {s_q q_\pa \over \qlsq(\qq)} \right] 
\delta(s\kpa - s_p p_\pa - s_q q_\pa)\, \delta_{k,pq} \, d{\bf p} \, d{\bf q} \, . 
$$
This equation tells us that we only have a nonlinear contribution when the wave polarities $s_p$ and $s_q$ 
are different. We recover here a well-known property of incompressible MHD: the nonlinear interactions are 
only due to counter-propagating Alfv\'en waves. This remark leads eventually to the following simplified form:
\be
\partial_t \qls (\kk) = \frac{\pi \, \epsilon^2}{2 \, b_0} \int \sum_{\Lambda_p, \Lambda_q \atop} 
\left( \frac{\sin \psi_k}{k} \right)^2 (\Lambda k + \Lambda_p p + \Lambda_q q)^2 
\left( \Lambda k - \Lambda_p p + \Lambda_q q \right)^2 
\label{kelim1}
\ee
$$
q_{\Lambda_p}^{- s} (\pp) \, \left[ q_{\Lambda_q}^s (\qq) - q_{\Lambda}^s (\kk) \right] 
\delta(p_\pa)\, \delta_{k,pq} \, d{\bf p} \, d{\bf q} \, . 
$$
This result is exactly the same as in \cite{galtier06} (see in particular Appendix D) where the MHD limit was 
discussed in the more general context of Hall MHD (the difference of a factor $8$ disappears after 
renormalization of the density tensor $q_{\Lambda}^s (\kk)$). Note that the comparison with \cite{galtier00} 
is not direct since the complex helicity basis was not used. 
The presence of $\delta(p_\pa)$ arises because of the three-wave frequency resonance condition. This means 
that in any triadic resonant interaction, there is always one wave that corresponds to a purely two-dimensional 
motion ($p_\pa=0$) whereas the two others have equal parallel components ($p_\pa=\kpa$). In other words, 
that means there is no nonlinear transfer along $\bbl$ and a cascade happens only in the perpendicular direction.

\subsection{Large-scale dynamics: inertial waves}

We consider the large-scale limit of (\ref{ke1}) for left-handed ($\Lambda=-s$) fluctuations. Then, we have 
at the leading order:
\ba
\alf &\to& -{s k d \over 2} \, , \\
\left( 2 + {\alfm}^2 {\alfmp}^2 {\alfmq}^2 - {\alfm}^2 - {\alfmp}^2 - {\alfmq}^2 \right)^2 &\to& 
\left( {64 \over k^2 p^2 q^2 d^6} \right)^2 \, , \\
\ols &\to& {2 \Omes s \kpa \over k} = \omega_I \, .
\ea
After introducing the previous expressions into (\ref{ke1}), we obtain:
\be
\partial_t \qls (\kk) = 
\frac{\pi \, \epsilon^2}{4b_0^2} \int \sum_{\Lambda_p, \Lambda_q \atop} 
\left( \frac{\sin \psi_k}{k} \right)^2 (\Lambda k + \Lambda_p p + \Lambda_q q)^2 
\frac{(\Lambda_q q - \Lambda_p p)^2}{p^2 q^2} {k^2 \omega_{\Lambda}^{-\Lambda} \over \kpa^2}
\label{kelim2}
\ee
$$
q_{\Lambda}^{-\Lambda} (\kk)
q_{\Lambda_p}^{-\Lambda_p} (\pp)
q_{\Lambda_q}^{-\Lambda_q} (\qq)
\left[ {k^2 \omega_{\Lambda}^{-\Lambda} \over q_{\Lambda}^{-\Lambda} (\kk)}
- {p^2 \omega_{\Lambda_p}^{-\Lambda_p} \over q_{\Lambda_p}^{-\Lambda_p} (\pp)} 
- {q^2 \omega_{\Lambda_q}^{-\Lambda_q} \over q_{\Lambda_q}^{-\Lambda_q} (\qq)} \right] 
\delta(\Omega_{k,pq})\, \delta_{k,pq} \, d{\bf p} \, d{\bf q} \, . 
$$
This result is exactly the same as in \cite{galtier03a} provided that the density tensor is correctly renormalized.

\subsection{Large-scale dynamics: magnetostrophic waves}
\label{emhd}

The last limit that we shall consider is the large-scale one for right-handed ($\Lambda=s$) fluctuations. 
We have at leading order:
\ba
\alf &\to& -{2s \over k d} \, , \\
\left( 2 + {\alfm}^2 {\alfmp}^2 {\alfmq}^2 - {\alfm}^2 - {\alfmp}^2 - {\alfmq}^2 \right)^2 &\to& 4 \, , \\
\ols &\to& {s \kpa kd b_0 \over 2} = \omega_M \, .
\ea
After introducing the previous expressions into (\ref{ke1}), we obtain:
\be
\partial_t \qls (\kk) = \frac{\pi \, \epsilon^2}{b_0^2} \int \sum_{\Lambda_p, \Lambda_q \atop} 
\left( \frac{\sin \psi_k}{k} \right)^2 (\Lambda k + \Lambda_p p + \Lambda_q q)^2 
\left( \Lambda_p p - \Lambda_q q \right)^2 {\omega_{\Lambda}^{\Lambda} \over \kpa^2}
\label{kelim3}
\ee
$$
q_{\Lambda}^{\Lambda} (\kk) q_{\Lambda_p}^{\Lambda_p} (\pp) q_{\Lambda_q}^{\Lambda_q} (\qq)
\left[ {\omega_{\Lambda}^{\Lambda} \over q_{\Lambda}^{\Lambda} (\kk)} 
- {\omega_{\Lambda_p}^{\Lambda_p} \over q_{\Lambda_p}^{\Lambda_p} (\pp)} 
- {\omega_{\Lambda_q}^{\Lambda_q} \over q_{\Lambda_q}^{\Lambda_q} (\qq)} \right] 
\delta(\Omega_{k,pq})\, \delta_{k,pq} \, d{\bf p} \, d{\bf q} \, . 
$$
This system has never been analyzed before, however, it is similar to the electron MHD case \citep{galtier03b}.

\section{Exact solutions for the turbulent spectra} \label{sect5}

We shall derive the exact solutions of the weak turbulence equations in three different limits: the large and small wave number 
limits with in the latter case a distinction between right and left polarizations. For that, we need to write the expression of the spectral 
density $\qls (\kk)$ in terms of explicit quantities like the kinetic and magnetic energies, the cross- and magnetic helicities. We inverse 
the system $(E^u, E^b, H^c, H^m) (\qls)$ and obtain:
\be
\qls (\kk) = 
\label{qexpand}
\ee
$$
{1 \over 2 \, ({\alf}^2-{\alfm}^2)}
\left[ {\alf}^2E^b(\kk) - E^u(\kk) + (\alf+\alfm) H^c(\kk) + \Lambda ({\alf}^2 -1) k H^m (\kk) \right] \, . 
$$
The introduction of expression (\ref{qexpand}) into (\ref{ke1}) leads to weak turbulence equations for $E^u$, $E^u$, $H^c$ and 
$H^m$. However, since we are only interested by three asymptotic limits (Alfv\'en, inertial and magnetostrophic wave turbulence)
for which we are able to derive the solutions, we may simplify the problem by taking the asymptotic values of the coefficients 
$\alf$ (see Section \ref{sect4}).

\subsection{Solutions for Alfv\'en wave turbulence}
\label{alfven1}

The small-scale limit of Alfv\'en wave turbulence is very well-known and has been analyzed in detail by \cite{galtier00}. For an 
application to the dynamo it is not the most relevant limit since the magnetic energy is expected to be accumulated at the 
largest scales of the system. Therefore, we will not give details about this regime but only recall the main properties. 
In the small-scale limit ($kd \to + \infty$), for which terms like ${\alfm}^2$ tend to $1$, an equipartition between the kinetic and 
magnetic energies is obtained and their dynamical equations tend to be identical. If we neglect the helicity contributions, the 
equation for the total energy gets reduce (see the derivation given in \cite{galtier06} where the helicity decomposition is used) 
and it is then possible to demonstrate that the axisymmetric bi-dimensional total energy spectrum follows the universal solution:
\be
E(\kpn,\kpa) \sim \kpn^{-2} f(\kpa) \, , 
\ee
where $f$ is an arbitrary function which traduces the dynamical decoupling of parallel planes in Fourier space. In other words, 
in Alfv\'en wave turbulence the cascade towards small-scales only happens in the perpendicular direction. This regime with its 
predictions has been observed in direct numerical simulations \citep{perez08,bigot08,bigot11}.

\subsection{Solutions for inertial wave turbulence}
\label{inertial1}

When the small-scale limit is taken with only the left polarization retained, one arrives to the inertial wave turbulence regime which was 
derived analytically by \cite{galtier03a} and studied numerically by \cite{Bellet}. Since $\alf \to 0$, we see immediately from relation 
(\ref{qexpand}) that the magnetic energy becomes negligible compared to the kinetic energy. Additionally, a simple analysis of 
equation (\ref{kelim2}) allows us to conclude that this turbulence becomes anisotropic. Indeed, if we assume that the nonlinear 
transfer is mainly the result of local interactions (\ie equilateral triads $k \approx p \approx q$), then the resonance condition 
(\ref{resonance}) simplifies to: 
\be
{s_p -s \over s s_p \qpa} \approx {s_p - s_q \over s_p s_q \kpa} \approx {s - s_q \over s s_q \ppa} \, . 
\ee
From equations (\ref{kelim2}), we see that only the interactions between two waves ($\pp$ and $\qq$) with opposite polarities 
($s=s_p=-s_q$ or $s=-s_p=s_q$; with $s=-\Lambda$) will contribute significantly to the nonlinear dynamics. It implies that either 
$\qpa \approx 0$ or $\ppa \approx 0$ which means that only a small transfer is allowed along $\Ome$. In other words, the local 
nonlinear interactions lead to anisotropic turbulence where small-scales are preferentially generated perpendicularly to the external 
rotation axis. Note that this approximation is particularly well verified initially if the turbulence is mainly excited in a limited band of scales: 
then, by nature the nonlinear interactions will be local and  will produce anisotropy. This short analysis allows us to consider the 
anisotropic limit of equation (\ref{kelim2}) for which $\kpn \gg \kpa$. We obtain the following equations: 
\be
{\partial \over \partial t} {E_k \brace H_k} = 
\label{iwt}
\ee
$$
{\Omega^2 \epsilon^2 \over 4} \sum_{s s_p s_q} \int 
{s \kpa s_p \ppa \over \kpn^2 \ppn^2 \qpn^2} \left( {s_q \qpn - s_p \ppn \over \omega_k} \right)^2 
(s \kpn + s_p \ppn + s_q \qpn)^2 \sin \theta_q 
$$

$$
{ E_q ( \ppn E_k - \kpn E_p ) + ( \ppn s H_k / \kpn - \kpn s_p H_p / \ppn ) s_q H_q / \qpn
\brace 
s \kpn \left[ E_q ( \ppn s H_k / \kpn - \kpn s_p H_p / \ppn) + ( \ppn E_k - \kpn E_p ) s_q H_q / \qpn \right] } 
$$

$$
\delta(s\omega_k + s_p\omega_p + s_q\omega_q) \, \delta(\kpa + \ppa + \qpa) \, \diso \, ,
$$
where $E_k \equiv E^u(\kpn,\kpa)$ and $H_k \equiv H^k(\kpn,\kpa)$ are respectively the axisymmetric bi-dimensional kinetic energy 
and kinetic helicity spectra, $\theta_q$ is the angle between the perpendicular wave vectors $\kkp$ and ${\bf \ppn}$ in the triangle 
made with ($\kkp$, ${\bf \ppn}$, ${\bf \qpn}$) and $\omega_k \simeq 2 \Omega \kpa / \kpn$. In equation (\ref{iwt}) the integration over 
perpendicular wave numbers is such that the triangular relation, $\kkp + {\bf \ppn} +{\bf \qpn}={\bf 0}$, must be satisfied. 
The exact solutions of equations (\ref{iwt}) were derived initially for a positive and constant kinetic energy flux \citep{galtier03a}; 
they read:
\ba
E_k &\sim& \kpn^{-5/2} \vert  \kpa \vert ^{-1/2} \, , \label{KZK1} \\
H_k &\sim& \kpn^{-3/2} \vert \kpa \vert^{-1/2} .
\label{KZK2}
\ea
In a situation where the turbulence is dominated by a (forward) helicity flux, it is necessary to consider the equation for 
the kinetic helicity to derive the other exact power law solutions. If we seek stationary solutions in the power law form 
$E_k \sim \kpn^{n} \vert \kpa\vert ^{m}$ and $H_k \sim \kpn^{\tilde n} \vert \kpa\vert ^{\tilde m}$, then the constant 
helicity flux solutions are more general and read \citep{galtier2014}: 
\ba
n+\nnn &=& -4 \, , \label{sol1} \\
m + \mmm &=& -1 \, . \label{sol2}
\ea
These solutions correspond to a positive helicity flux and thus a direct cascade. 
The cascade along the rotation axis being strongly reduced, the most important scaling law is therefore the one 
for the perpendicular wave numbers. It is remarkable to see that the exact solution (\ref{sol1}) corresponds to the 
empirical law observed in a myriad of direct numerical simulations where the helicity transfer dominates the energy transfer 
\citep[see \eg][]{mininni09,mininni12}. The domain of convergence of this family of solutions writes:
\ba
-3 < n + m  < -2 \, , \label{d1} \\
-2 < \nnn + \mmm < -1 \, . \label{d2} 
\ea
The spectral solutions of the inertial wave turbulence regime are at the border line of the domain of convergence. However, 
since the problem is strongly anisotropic and the inertial range in the parallel direction is strongly reduced with a cascade 
almost only in the perpendicular direction, we may neglect the inertial range in the parallel direction which is equivalent to say 
$m=\mmm=0$. Then, we obtain a classical result of weak turbulence in the sense that the power law indices of the exact solutions 
(\ref{KZK1})--(\ref{KZK2}) fall exactly at the middle of the domains of locality (\ref{d1})--(\ref{d2}). In conclusion, we see that the 
turbulent spectra does not correspond necessarily to the so-called maximal helicity state which is a particular solution of the 
Schwarz inequality $H(\kk) \le k E(\kk)$ (here we consider directly the weak turbulence limit for which the polarization term 
\citep{Cambon89} does not contribute) and for which $n=\nnn-1=-5/2$. As the helicity transfer increases the power law indices 
$n$ and $\nnn$ get closer. The condition of locality gives, however, a limit to this convergence namely $n=\nnn=-2$.

\subsection{Solutions for magnetostrophic wave turbulence}
\label{MI1}

The small-scale limit of expression (\ref{ke1}) can lead to the magnetostrophic wave turbulence equations if only the right polarization 
is retained. As for inertial wave turbulence, we may show from equation (\ref{kelim2}) that this turbulence becomes naturally anisotropic. 
Indeed, is we consider that the nonlinear transfer is mainly due to local interactions ($k \approx p \approx q$), the resonance condition 
(\ref{resonance}) simplifies to: 
\be
{s_p -s \over \qpa} \approx {s_p - s_q \over \kpa} \approx {s - s_q \over \ppa} \, . 
\ee
From equations (\ref{kelim3}), we see that only the interactions between two waves ($\pp$ and $\qq$) with opposite polarities 
($s=s_p=-s_q$ or $s=-s_p=s_q$; with $s=\Lambda$) will contribute significantly to the nonlinear dynamics. It implies that either 
$\qpa \approx 0$ or $\ppa \approx 0$ which means that only a small transfer is allowed along $\Ome$. As for inertial wave turbulence,
(i) the local nonlinear interactions lead to anisotropic turbulence where the cascade is preferentially generated perpendicularly to the 
external rotation axis, and (ii) the approximation is particularly well verified initially if the turbulence is mainly excited in a limited band 
of scales since then, by nature the nonlinear interactions will be local. 
From this discussion, it seems relevant to take the anisotropic limit ($\kpn \gg \kpa$) of equation (\ref{kelim3}) which gives: 
\be
\partial_t {E_k \brace H_k} =
\label{isom}
\ee
$$ 
{\epsilon^2 \over 16} \, \sum_{s s_p s_q} \int \, {s_p \, \ppn \kpa \ppa \over \qpn} 
\left({s_q \qpn - s_p \ppn \over \kpa} \right)^2 (s \kpn + s_p \ppn + s_q \qpn)^2 \, \sin \theta_q
$$

$${ s \kpn \left[ E_q ( \ppn E_k - \kpn E_p )/ (\kpn \ppn \qpn) + \, s_q \, H_q \, ( s H_k - s_p H_p ) \right]
\brace
E_q ( s \, H_k - s_p H_p)/\qpn + s_q \, H_q ( \ppn E_k - \kpn E_p )/(\kpn \ppn)} 
$$

$$
\delta(\kpa + \ppa + \qpa) \, \delta(s\kpn\kpa + s_p\ppn\ppa + s_q\qpn\qpa) \, \diso \, ,
$$
where $E_k \equiv E^b(\kpn,\kpa)$ and $H_k \equiv H^m(\kpn,\kpa)$ are respectively the axisymmetric bi-dimensional magnetic 
energy and magnetic helicity spectra and, as before, $\theta_q$ is the angle between the perpendicular wave vectors $\kkp$ and 
${\bf \ppn}$ in the triangle made with ($\kkp$, ${\bf \ppn}$, ${\bf \qpn}$). In equation (\ref{isom}) the integration over perpendicular 
wave numbers is such that the triangular relation, $\kkp + {\bf \ppn} +{\bf \qpn}={\bf 0}$, must be satisfied. 
To derive the exact solutions, we have to introduce the following power law forms for the spectra 
$E_k \sim \kpn^{n} \vert \kpa\vert ^{m}$ and $H_k \sim \kpn^{\tilde n} \vert \kpa\vert ^{\tilde m}$, and apply a bi-homogeneous conformal 
transform \citep{ZLF92,nazarenko11} which consists in doing the following manipulation on the wave numbers $\ppn$, $\qpn$, $\ppa$ 
and $\qpa$:
\be
\begin{array}{lll}
\ppn &\to & \kpn^2 / \ppn \, , \\[.2cm]
\qpn &\to & \kpn \qpn / \ppn \, , \\[.2cm]
\vert \ppa \vert &\to & \kpa^2 / \vert \ppa\vert  \, , \\[.2cm]
\vert \qpa \vert &\to & \vert \kpa\vert \vert  \qpa\vert  / \vert \ppa \vert  \, . 
\end{array}
\label{trans}
\ee
This exercise for the energy equation gives the positive and constant energy flux solutions:
\ba
E_k &\sim& \kpn^{-5/2} \vert  \kpa \vert ^{-1/2} \, , \label{KZK1bb} \\
H_k &\sim& \kpn^{-7/2} \vert \kpa \vert^{-1/2} .
\label{KZK2bb}
\ea
The same transform applied to the helicity equation extends the previous solutions to a family of solutions: 
\ba
n+\nnn &=& -6 \, , \label{sol1b} \\
m + \mmm &=& -1 \, . \label{sol2b}
\ea
This family of solutions corresponds to a negative and constant magnetic helicity flux, hence the possible existence of an inverse
cascade of helicity and the accumulation of magnetic energy at large-scales. 
Since the cascade along the uniform magnetic field is strongly reduced, the most important scaling law is therefore the one for the 
perpendicular wave numbers. The domain of convergence of these solutions writes: 
\ba
-3 < n + m  < -2 \, , \label{dd1} \\
-4 < \nnn + \mmm < -3 \, . \label{dd2} 
\ea
We see that with the previous solutions (obtained from the energy or the helicity equations) we are at the border line
of the domain of convergence. However, we also know that this problem is strongly anisotropic and the inertial range 
in the parallel direction is strongly reduced with a cascade almost only in the perpendicular direction. Actually, if we neglect 
the inertial range in the parallel direction (which is equivalent to say $m=\mmm=0$) we obtain again -- like for the inertial wave
turbulence regime -- a classical result of weak turbulence in the sense that the power law indices of the exact solutions 
(\ref{KZK1bb})--(\ref{KZK2bb}) fall exactly at the middle of the domains of locality (\ref{dd1})--(\ref{dd2}). Note that the solutions 
found do not allow a crossing of the spectra since the case $n = \tilde n = -3$ appears as an asymptotic limit. Note also that  
the classical phenomenology presented in Section \ref{sect4bis} gives the particular asymptotic solution $\tilde n = -3$. 
It is only through a deep mathematical treatment that this family of solutions may be discovered. This situation is also found 
for the inertial wave turbulence regime for which many papers have been devoted but where no consistent anisotropic 
phenomenology has been proposed. For that reason, these exact solutions may be qualified as highly non-trivial. 
Finally, it is interesting to remark that the process of inverse cascade described here is limited in scales since the basic 
assumption made for the analysis is that $k_\perp \gg k_\parallel$. When this condition is broken (with \eg $k_\perp \ll k_\parallel$), 
the previous local analysis made on the resonance condition becomes irrelevant and the theoretical predictions not possible.

\section{Discussion} \label{sect7}

In this paper I have developed a weak turbulence theory for rotating MHD under the presence of a parallel uniform magnetic field. 
The theory is expected to be relevant for the magnetostrophic dynamo with applications to Earth and giant planets for which 
a small ($\sim 10^{-6}$) Rossby number is expected. An important question which may be investigated is the mechanism of 
regeneration of a large-scale magnetic field through an inverse cascade of hybrid helicity. 
A key length scale in this problem is the magneto-inertial length $d$ which indicates the basin of attraction for the dynamics. 
Basically, if the scales considered are larger than $d$ (in other words if $kd < 1$), we fall in the inertial or magnetostrophic wave 
turbulence regime, the precise localization being determined by the nature of the polarization (left or right respectively). If, 
however, the scales are smaller than $d$ ($kd > 1$) and if the condition for weak turbulence are still satisfied (with a wave period 
much smaller than the eddy-turn-over time, otherwise the turbulence is strong), then we fall in the Alfv\'en wave turbulence regime. 
It is interesting to note that the magnetostrophic regime -- also called strong-field regime -- is driven by a nonlinear equation 
(\ref{magn3}) similar to a well-known system in plasma physics called electron MHD \citep{Kingsep} which finds applications, 
\eg in space plasmas \citep{galtier06b}. 

\begin{figure}
\centerline{\includegraphics[width=11cm]{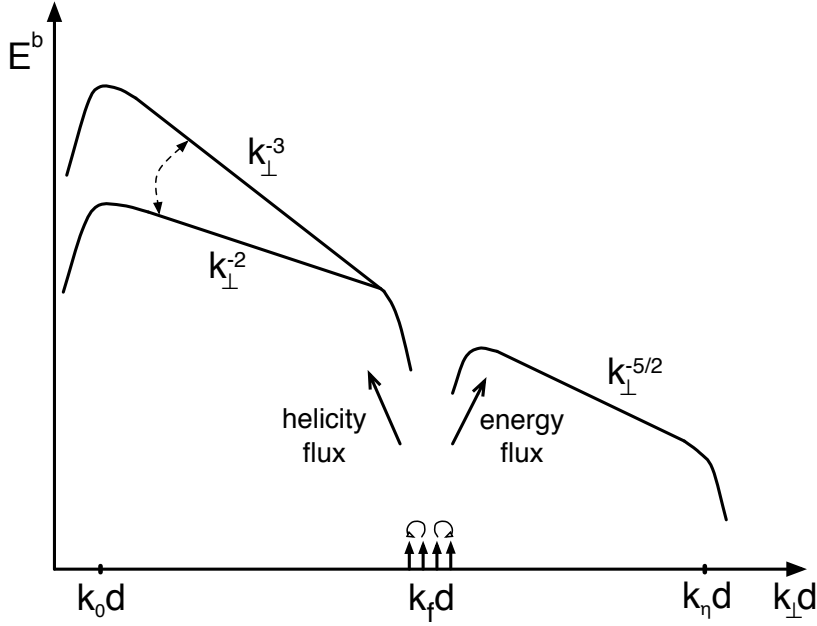}}
\caption{Magnetic energy spectrum in the weak magnetostrophic turbulence regime ($kd<1$) and with a (\eg convective) 
forcing applied in a range of intermediate scales $k_f$. While the direct energy cascade gives a unique scaling, the inverse helicity 
cascade may lead to a family of solutions confined between $\kpn^{-2}$ and $\kpn^{-3}$ due to the entanglement of the helicity 
and energy. In practice, the inertial ranges are limited by the largest scale of the system $k_0$ (\eg the size of the outer core) and 
the dissipative scale (\eg the magnetic one) $k_\eta$.}
\label{fig4}
\end{figure}

By using a complex helicity decomposition, the asymptotic weak turbulence equations have been derived which describe 
the long-time behavior of weakly dispersive interacting waves {\it via} three-wave processes. For magnetostrophic 
wave turbulence, the theory predicts that the magnetic energy is asymptotically larger than the kinetic energy when one goes 
to large-scales, whereas it is the inverse for inertial wave turbulence. The analysis of the resonance conditions has been used 
to prove the anisotropic nature of the nonlinear transfer with a stronger cascade perpendicular than parallel to the rotating axis. 
Then, the reduced forms of the general equations of weak turbulence have been obtained in the three relevant limits discussed 
above with their exact power law solutions after the application of the Kuznetsov--Zakharov transform (see figure \ref{fig4}). 
The large-scale (magnetostrophic and inertial) solutions can be highly non-trivial in the sense that the classical anisotropic 
phenomenology is only able to catch the correct scaling for the constant energy flux solutions which are dimensionally 
compatible with a maximal helicity state. The solutions for the constant (magnetic or kinetic) helicity flux are, however, not recovered 
with a consistent phenomenology. The non triviality resides in an entanglement relation which implies the energy and helicity spectra 
power law indices. At large-scales ($kd < 1$), whereas a direct cascade of kinetic helicity is expected which is well observed in direct 
numerical simulations of pure rotating hydrodynamic turbulence \citep[see \eg][]{mininni09,mininni12}, an inverse cascade of 
magnetic helicity is predicted. Since the magnetostrophic wave turbulence regime is similar to the electron MHD one where 
an inverse cascade has already been observed in direct numerical simulations \citep{shaikh05,cho11} 
we may think that it is a reasonable prediction. Then, in the context of the dynamo problem the main 
question is: at which scale $k_f$ the system is driven\,? Indeed, if the forcing scales is such that $k_f d < 1$ we fall in the large-scale
regime (magnetostrophic basin of attraction; see figure \ref{fig4}) and the dynamo mechanism may happen through an inverse 
cascade of hybrid helicity which is dominated by the magnetic helicity. However, if that scale is such that $k_f d > 1$, then we fall 
in the small-scale regime (Alfv\'en basin of attraction) and the regeneration of the magnetic field becomes more difficult since the 
hybrid helicity is dominated by the cross-helicity which cascades in the forward (to small-scales) direction \citep{galtier00}. 
It is important to recall that the magnetic helicity is {\it not} an inviscid invariant in the weak (non rotating) MHD turbulence regime 
where a uniform magnetic field is present; the question of the regeneration of a large-scale magnetic field needs therefore a new 
ingredient like the Coriolis force to be relevant. 

The present theory may be useful to better understand the magnetostrophic dynamo with applications to Earth and giant planets. 
Although our theory is a crude model for such a problem (for example, we assume a magnetic Reynolds numbers large enough 
for the development of an extended inertial range and we do not include the geometry effects with boundary conditions), it is believed 
that the dynamics obtained here at asymptotically small Rossby number opens new perspectives. For example, in the case of the 
outer core of Earth a rough evaluation of the magneto-inertial length gives $d \approx 1$km \citep{Finlay10}. If we consider that the 
forcing due to convection has a typical length scale of $1/k_f \approx 100$km then the conditions for an inverse cascade 
are satisfied. Another question is about the surprising axisymmetry of planets like Earth, Jupiter or Saturn where the rotation and 
magnetic axes are close and even almost perfect for Saturn. The present turbulence theory gives a possible answer. Indeed, the 
rotating MHD equations in presence of a uniform magnetic field have in general only one inviscid invariant, the total energy. 
It is only when the rotation and magnetic axes
are aligned that a second inviscid invariant appears, namely the hybrid helicity. It is precisely this second invariant which can generate 
a turbulent dynamo through an inverse cascade. We may believe that as long as the angle $\theta$ between $\Ome$ and $\bbl$ 
remains reasonably small the inverse cascade may still operate. According to this remark, it is not surprising that a strong alignment, 
with $\theta \le 10^{\rm o}$, is generally observed for the previous magnetized planets. 
The initial phase of the dynamo has not been discussed until now but it deserves a short discussion. Since in absence of a 
uniform magnetic field the magnetic helicity is an inviscid invariant of rotating MHD, an inverse cascade may happen. This 
mechanism is, however, under the influence of the Coriolis force which renders the dynamics anisotropic. Then, we may expect 
the generation of a large-scale magnetic preferentially aligned with the rotation axis. After this initial phase, it seems then natural to 
consider the regime described in the present paper.

\appendix
\section{Useful relationships}
\label{relation}

From the quantity:
\be
\xi_{\Lambda}^s = \frac{- s kd}{\left( - s \Lambda + \sqrt{1 + k^2 d^2} \right)} \, , 
\ee
it is possible to derive the following useful identities: 
\ba
\alf \, \alfm &=& -1 \, , \\
\xi_{-\Lambda}^{-s} &=& - \alf \, , \label{ident5} \\
\alf + \alfm &=& - {2 \over \Lambda kd} \, , \\
\alf - \alfm &=& - {2s \over kd} \sqrt{1+k^2 d^2} \, , \label{A13} \\
1-{\alf}^2 &=& {2 \Lambda \Omes \over b_0 k} \, \alf \, . 
\ea
We also have the remarkable relations:
\ba
\omega^s_s \omega^s_{-s} &=& (\kpa \bbl)^2 \, , \\
{\omega^s_{s}}^2 &\le& (\kpa \bbl)^2 \le {\omega^s_{-s}}^2 \, .
\ea

\section{Helicity decomposition}
\label{hdecomp}

The projection of the Fourier transform of the original vectors $\uu ({\bf x})$ and $\bb ({\bf x})$ on the helicity basis gives:
\ba
\uu_\kk &=& \sum_{\Lambda} \, {\cal U}_{\Lambda} (\kk) \, {\bf h^{\Lambda}_k}  \, , \\
\bb_\kk &=& \sum_{\Lambda} \, {\cal B}_{\Lambda} (\kk) \, {\bf h^{\Lambda}_k}  \, .
\ea
If we inverse the system, we find the following relations for the velocity components: 

\ba
{\cal U}_+ (\kk) &=& {1 \over 2 k \kpn} \left[ k_x \kpa u_x + k_y \kpa u_y - \kpn^2 u_z + i k (k_y u_x - k_x u_y)\right] \, ,\\
{\cal U}_- (\kk) &=& {1 \over 2 k \kpn} \left[ k_x \kpa u_x + k_y \kpa u_y - \kpn^2 u_z - i k (k_y u_x - k_x u_y) \right] \, .
\ea
Similar relations are found for the magnetic field. Note that such helicity decomposition cannot be applied for the modes $\kpn=0$.

\section{Derivation of the weak turbulence equations}
\label{derivation}

The starting point of the derivation of the weak turbulence equations is the fundamental equation 
(\ref{fonda2}). We write successively equations for the second and third-order moments:
\be
\partial_t \langle \aak \aakp \rangle = 
\label{second}
\ee
$$
\frac{\epsilon \, d^2}{16}
\int \sum_{\Lambda_p, \Lambda_q \atop s_p, s_q}
\frac{\alfmq - \alfmp}{\alf - \alfm} \, 
M{{\Lambda \Lambda_p \Lambda_q \atop s \, s_p \, s_q} \atop -k \, p \, q}
\, \langle \aap \aaq \aakp \rangle \, e^{-i \Omega_{pq,k} t} \, \delta_{pq,k} \, 
d{\bf p} \, d{\bf q}
$$
$$
+
$$
$$
\frac{\epsilon \, d^2}{16}
\int \sum_{\Lambda_p, \Lambda_q \atop s_p, s_q}
\frac{\alfmq - \alfmp}{\alfprime - \alfmprime} \, 
M{{\Lambda^{\prime} \Lambda_p \Lambda_q \atop s^{\prime} \, s_p \, s_q} 
\atop -k^{\prime} \, p \, q}
\, \langle \aap \aaq \aak \rangle \, e^{-i \Omega_{pq,k^{\prime}} t} \, 
\delta_{pq,k^{\prime}} \, d{\bf p} \, d{\bf q} \, ,
$$
and:
\be
\partial_t \langle \aak \aakp \aakdp \rangle = 
\label{third}
\ee
$$
\frac{\epsilon \, d^2}{16}
\int \sum_{\Lambda_p, \Lambda_q \atop s_p, s_q}
\frac{\alfmq - \alfmp}{\alf - \alfm} \, 
M{{\Lambda \Lambda_p \Lambda_q \atop s \, s_p \, s_q} \atop -k \, p \, q}
\, \langle \aap \aaq \aakp \aakdp \rangle \, e^{-i \Omega_{pq,k} t} \, \delta_{pq,k} \, 
d{\bf p} \, d{\bf q}
$$
$$
+
$$
$$
\frac{\epsilon \, d^2}{16}
\int \sum_{\Lambda_p, \Lambda_q \atop s_p, s_q}
\frac{\alfmq - \alfmp}{\alfprime - \alfmprime} \, 
M{{\Lambda^{\prime} \Lambda_p \Lambda_q \atop s^{\prime} \, s_p \, s_q} \atop 
-k^{\prime} \, p \, q}
\, \langle \aap \aaq \aak \aakdp \rangle \, e^{-i \Omega_{pq,k^{\prime}} t} \, 
\delta_{pq,k^{\prime}} \, d{\bf p} \, d{\bf q}
$$
$$
+
$$
$$
\frac{\epsilon \, d^2}{16}
\int \sum_{\Lambda_p, \Lambda_q \atop s_p, s_q}
\frac{\alfmq - \alfmp}{\alfdprime - \alfmdprime} \, 
M{{\Lambda^{\prime \prime} \Lambda_p \Lambda_q \atop s^{\prime \prime} \, s_p \, s_q} 
\atop -k^{\prime \prime} \, p \, q} \, \langle \aap \aaq \aak \aakp \rangle \, 
e^{-i \Omega_{pq,k^{\prime \prime}} t} \, \delta_{pq,k^{\prime \prime}} \, d{\bf p} \, 
d{\bf q}\, .
$$
We shall write an asymptotic closure \citep{nazarenko11} for our system. For that, we basically need to write 
the fourth-order moment in terms of a sum of the fourth-order cumulant plus products of second order ones. 
The asymptotic closure depends on two ingredients: the first is the degree to which the linear waves interact 
to randomize phases; the second relies on the fact that the nonlinear regeneration of the third-order moment 
by the fourth-order moment in equation (\ref{third}) depends more on the product of the second order 
moments than it does on the fourth order cumulant. The fourth--order moment decomposes into the sum of 
three products of second--order moments, and a fourth--order cumulant. The latter does not contribute to 
secular behavior, and among the other products one is absent because of the homogeneity assumption. 
If we use the symmetric relations (\ref{prop1})--(\ref{prop4}) and perform wavevector integrations, summations 
over polarities and time integration, then equation (\ref{third}) becomes:
\be
\langle \aak \aakp \aakdp \rangle = 
\frac{\epsilon \, d^2}{16} \, \Delta(\Omega_{k k^{\prime} k^{\prime \prime}}) \, 
\delta_{k k^{\prime} k^{\prime \prime}}
\label{third2}
\ee
$$
\{ \, 
\left[\frac{\alfmdprime - \alfmprime}{\alf - \alfm} 
\left(M{{\Lambda \Lambda^{\prime} \Lambda^{\prime \prime} \atop s \, s^{\prime} \, 
s^{\prime \prime}} \atop k \, k^{\prime} \, k^{\prime \prime}}\right)^* + 
\frac{\alfmprime - \alfmdprime}{\alf - \alfm} 
\left(M{{\Lambda \Lambda^{\prime \prime} \Lambda^{\prime} \atop 
s \, s^{\prime \prime} \, s^{\prime}} \atop k \, k^{\prime \prime} \, k^{\prime}}\right)^*
\right] \qlsprime \, \qlsdprime
$$
$$
+
$$
$$
\left[\frac{\alfmdprime - \alfm}
{\alfprime - \alfmprime} 
\left(M{{\Lambda^{\prime} \Lambda \Lambda^{\prime \prime} \atop s^{\prime} \, s \, 
s^{\prime \prime}} \atop k^{\prime} \, k \, k^{\prime \prime}}\right)^* + 
\frac{\alfm - \alfmdprime}{\alfprime - \alfmprime} 
\left(M{{\Lambda^{\prime} \Lambda^{\prime \prime} \Lambda \atop 
s^{\prime} \, s^{\prime \prime} \, s} \atop k^{\prime} \, k^{\prime \prime} \, k}\right)^*
\right] \qls \, \qlsdprime
$$
$$
+
$$
$$
\left[\frac{\alfm - \alfmprime}{\alfdprime - \alfmdprime} 
\left(M{{\Lambda^{\prime \prime} \Lambda^{\prime} \Lambda 
\atop s^{\prime \prime} \, s^{\prime} \, s} 
\atop k^{\prime \prime} \, k^{\prime} \, k}\right)^* + 
\frac{\alfmprime - \alfm}{\alfdprime - \alfmdprime} 
\left(M{{\Lambda^{\prime \prime} \Lambda \Lambda^{\prime} \atop 
s^{\prime \prime} \, s \, s^{\prime}} \atop k^{\prime \prime} \, k \, k^{\prime}}\right)^*
\right] \qlsprime \, \qls \, \} \, ,
$$
where:
\be
\Delta(\Omega_{k k^{\prime} k^{\prime \prime}})= 
\int_0^t e^{i \Omega_{k k^{\prime} k^{\prime \prime}} t^{\prime}} dt^{\prime} 
= {e^{i \Omega_{k k^{\prime} k^{\prime \prime}}t} - 1 
\over i \Omega_{k k^{\prime} k^{\prime \prime}}} \, .
\ee
The introduction of symmetric relations (\ref{prop1})--(\ref{prop4}) into (\ref{third2}) allows us to simplify 
further the previous equation; one obtains:
\be
\langle \aak \aakp \aakdp \rangle = 
\frac{\epsilon \, d^2}{8} \, \Delta(\Omega_{k k^{\prime} k^{\prime \prime}}) \, 
\delta_{k k^{\prime} k^{\prime \prime}}
\left(M{{\Lambda \Lambda^{\prime} \Lambda^{\prime \prime} \atop s \, s^{\prime} \, 
s^{\prime \prime}} \atop k \, k^{\prime} \, k^{\prime \prime}}\right)^*
\label{third3}
\ee
$$
\left[
\frac{\alfmdprime - \alfmprime}{\alf - \alfm} \qlsprime \, \qlsdprime + 
\frac{\alfm - \alfmdprime}{\alfprime - \alfmprime} \qls \, \qlsdprime + 
\frac{\alfmprime - \alfm}{\alfdprime - \alfmdprime} \qls \, \qlsprime
\right] \, .
$$
We insert expression (\ref{third3}) into equation (\ref{second}); it leads to: 
\be
\partial_t \qls (\kk) = 
\label{secondbis}
\ee
$$
\frac{\epsilon^2 \, d^4}{128}
\int \sum_{\Lambda_p, \Lambda_q \atop s_p, s_q}
\frac{\alfmq - \alfmp}{\alf - \alfm} \, 
\left|M{{\Lambda \Lambda_p \Lambda_q \atop s \, s_p \, s_q} \atop -k \, p \, q}\right|^2
\Delta(\Omega_{p q, k}) \, e^{- i \Omega_{p q, k} t} \, \delta_{pq,k}
$$
$$
\left[
\frac{\alfm - \alfmq}{\alfp - \alfmp} \qls \, \qlsq + 
\frac{\alfmp - \alfm}{\alfq - \alfmq} \qls \, \qlsp + 
\frac{\alfmq - \alfmp}{\alf - \alfm} \qlsp \, \qlsq
\right] 
d{\bf p} \, d{\bf q}
$$
$$
+
$$
$$
\frac{\epsilon^2 \, d^4}{128}
\int \sum_{\Lambda_p, \Lambda_q \atop s_p, s_q}
\frac{\alfmq - \alfmp}{\alfprime - \alfmprime} \, 
\left|M{{\Lambda^{\prime} \Lambda_p \Lambda_q \atop s^{\prime} \, s_p \, s_q} 
\atop -k^{\prime} \, p \, q}\right|^2
\Delta(\Omega_{p q, k^{\prime}}) \, e^{- i \Omega_{p q, k^{\prime}} t} \, 
\delta_{pq,k^{\prime}}
$$
$$
\left[
\frac{\alfmprime - \alfmq}{\alfp - \alfmp} \qlsprime \, \qlsq + 
\frac{\alfmp - \alfmprime}{\alfq - \alfmq} \qlsprime \, \qlsp + 
\frac{\alfmq - \alfmp}{\alfprime - \alfmprime} \qlsp \, \qlsq
\right] 
d{\bf p} \, d{\bf q} \, .
$$
The long-time behavior of the weak turbulence equation (\ref{secondbis}) is given by the Riemman-Lebesgue 
Lemma which tells us that, for $t \to +\infty$, we have:
\be
e^{-ix t} \Delta(x) = \Delta(-x) \to \pi \delta(x) - i {\cal P}(1/x) \, ,
\ee
where ${\cal P}$ is the principal value of the integral. The two terms of equation (\ref{secondbis}) are complex 
conjugated therefore if in the second term we replace the dummy integration variables $\pp$, $\qq$, by $-\pp$, 
$-\qq$, we can simplify further equation (\ref{secondbis}) since, in particular, principal value terms compensate 
exactly. Finally, we obtain the weak turbulence equation:
\be
\partial_t \qls (\kk) = 
\label{ke0}
\ee
$$
\frac{\pi \, \epsilon^2 \, d^4}{64}
\int \sum_{\Lambda_p, \Lambda_q \atop s_p, s_q} \frac{\alfmq - \alfmp}{\alf - \alfm} \, 
\left|M{{\Lambda \Lambda_p \Lambda_q \atop s \, s_p \, s_q} \atop -k \, p \, q}\right|^2
\delta(\Omega_{k, p q}) \, \delta_{k,pq}
$$
$$
\left[
\frac{\alfm - \alfmq}{\alfp - \alfmp} \, \qls \, \qlsq + 
\frac{\alfmp - \alfm}{\alfq - \alfmq} \, \qls \, \qlsp + 
\frac{\alfmq - \alfmp}{\alf - \alfm} \, \qlsp \, \qlsq
\right] 
d{\bf p} \, d{\bf q} \, , 
$$
where:
$$
\left|M{{\Lambda \Lambda_p \Lambda_q \atop s \, s_p \, s_q} \atop -k \, p \, q}\right|^2 =
\left( \frac{\sin \psi_k}{k} \right)^2 k^2 p^2 q^2 \, (\Lambda k + \Lambda_p p + \Lambda_q q)^2 \, 
$$
$$
{\alf}^2{\alfp}^2{\alfq}^2
\left( 2 + {\alfm}^2 {\alfmp}^2 {\alfmq}^2 - {\alfm}^2 - {\alfmp}^2 - {\alfmq}^2 \right)^2 \, 
\, .
$$
The last step that we have to follow to obtain the same expression as (\ref{ke1}) is to include the resonance 
relations (\ref{resonance}) into the previous equations.

\bibliographystyle{jfm}
\bibliography{liste}

\begin{thebibliography}{89}
\expandafter\ifx\csname natexlab\endcsname\relax\def\natexlab#1{#1}\fi

\bibitem[{Baroud} {\em et~al.\/}(2002){Baroud}, {Plapp}, {She} \&
  {Swinney}]{baroud}
{\sc {Baroud}, C.~N., {Plapp}, B.~B., {She}, Z.-S. \& {Swinney}, H.~L.} 2002
  {Anomalous Self-Similarity in a Turbulent Rapidly Rotating Fluid}. {\em Phys.
  Rev. Lett.\/} {\bf 88}~(11), 114501.

\bibitem[{Bellet} {\em et~al.\/}(2006){Bellet}, {Godeferd}, {Scott} \&
  {Cambon}]{Bellet}
{\sc {Bellet}, F., {Godeferd}, F.~S., {Scott}, J.~F. \& {Cambon}, C.} 2006
  {Wave turbulence in rapidly rotating flows}. {\em J. Fluid Mech.\/} {\bf
  562}, 83--121.

\bibitem[{Benney} \& {Newell}(1967)]{BN67}
{\sc {Benney}, D.~J. \& {Newell}, A.~C.} 1967 {Sequential time closures for
  interacting random waves}. {\em J. Math. Phys.\/} {\bf 46}, 363--393.

\bibitem[{Benney} \& {Newell}(1969)]{benney69}
{\sc {Benney}, D.~J. \& {Newell}, A.~C.} 1969 {Random wave closures}. {\em
  Studies in Applied Math.\/} {\bf 48}, 29--53.

\bibitem[{Benney} \& {Saffman}(1966)]{benney66}
{\sc {Benney}, D.~J. \& {Saffman}, P.~G.} 1966 {Nonlinear Interactions of
  Random Waves in a Dispersive Medium}. {\em R. Soc. Lond. Proc. Series A\/}
  {\bf 289}, 301--320.

\bibitem[{Berhanu} {\em et~al.\/}(2007){Berhanu}, {Monchaux}, {Fauve},
  {Mordant}, {P{\'e}tr{\'e}lis}, {Chiffaudel}, {Daviaud}, {Dubrulle},
  {Mari{\'e}}, {Ravelet}, {Bourgoin}, {Odier}, {Pinton} \& {Volk}]{Berhanu}
{\sc {Berhanu}, M., {Monchaux}, R., {Fauve}, S., {Mordant}, N.,
  {P{\'e}tr{\'e}lis}, F., {Chiffaudel}, A., {Daviaud}, F., {Dubrulle}, B.,
  {Mari{\'e}}, L., {Ravelet}, F., {Bourgoin}, M., {Odier}, P., {Pinton}, J.-F.
  \& {Volk}, R.} 2007 {Magnetic field reversals in an experimental turbulent
  dynamo}. {\em Europhys. Lett.\/} {\bf 77}, 59001.

\bibitem[{Bigot} \& {Galtier}(2011)]{bigot11}
{\sc {Bigot}, B. \& {Galtier}, S.} 2011 {Two-dimensional state in driven
  magnetohydrodynamic turbulence}. {\em Phys. Rev. E\/} {\bf 83}~(2), 026405.

\bibitem[{Bigot} {\em et~al.\/}(2008){Bigot}, {Galtier} \& {Politano}]{bigot08}
{\sc {Bigot}, B., {Galtier}, S. \& {Politano}, H.} 2008 {Development of
  anisotropy in incompressible magnetohydrodynamic turbulence}. {\em Phys. Rev.
  E\/} {\bf 78}~(6), 066301.

\bibitem[{Bourouiba}(2008)]{bourouiba08}
{\sc {Bourouiba}, L.} 2008 {Discreteness and resolution effects in rapidly
  rotating turbulence}. {\em Phys. Rev. E\/} {\bf 78}~(5), 056309.

\bibitem[{Braginsky} \& {Roberts}(1995)]{braginsky}
{\sc {Braginsky}, S.~I. \& {Roberts}, P.~H.} 1995 {Equations governing
  convection in earth's core and the geodynamo}. {\em Geophys. Astrophys. Fluid
  Dyn.\/} {\bf 79}, 1--97.

\bibitem[{Brandenburg}(2001)]{brandenburg}
{\sc {Brandenburg}, A.} 2001 {The Inverse Cascade and Nonlinear Alpha-Effect in
  Simulations of Isotropic Helical Hydromagnetic Turbulence}. {\em Astrophys.
  J.\/} {\bf 550}, 824--840.

\bibitem[{Cambon} \& {Jacquin}(1989)]{Cambon89}
{\sc {Cambon}, C. \& {Jacquin}, L.} 1989 {Spectral approach to non-isotropic
  turbulence subjected to rotation}. {\em J. Fluid Mech.\/} {\bf 202},
  295--317.

\bibitem[{Cambon} {\em et~al.\/}(1997){Cambon}, {Mansour} \&
  {Godeferd}]{Cambon97}
{\sc {Cambon}, C., {Mansour}, N.~N. \& {Godeferd}, F.~S.} 1997 {Energy transfer
  in rotating turbulence}. {\em J. Fluid Mech.\/} {\bf 337}, 303--332.

\bibitem[{Chen} {\em et~al.\/}(2003{\natexlab{{\em a\/}}}){Chen}, {Chen} \&
  {Eyink}]{chen1}
{\sc {Chen}, Q., {Chen}, S. \& {Eyink}, G.~L.} 2003{\natexlab{{\em a\/}}} {The
  joint cascade of energy and helicity in three-dimensional turbulence}. {\em
  Phys. Fluids\/} {\bf 15}, 361--374.

\bibitem[{Chen} {\em et~al.\/}(2003{\natexlab{{\em b\/}}}){Chen}, {Chen},
  {Eyink} \& {Holm}]{chen2}
{\sc {Chen}, Q., {Chen}, S., {Eyink}, G.~L. \& {Holm}, D.~D.}
  2003{\natexlab{{\em b\/}}} {Intermittency in the Joint Cascade of Energy and
  Helicity}. {\em Phys. Rev. Lett.\/} {\bf 90}~(21), 214503.

\bibitem[{Cho}(2011)]{cho11}
{\sc {Cho}, J.} 2011 {Magnetic Helicity Conservation and Inverse Energy Cascade
  in Electron Magnetohydrodynamic Wave Packets}. {\em Phys. Rev. Lett.\/} {\bf
  106}~(19), 191104.

\bibitem[Craya(1954)]{craya58}
{\sc Craya, A.} 1954 {Contribution \`a l'analyse de la turbulence associ\'ee
  \`a des vitesses moyennes}. {\em P.S.T. Minist\`ere de l'Air\/} {\bf 345}.

\bibitem[{Davidson}(2004)]{davidsonbook}
{\sc {Davidson}, P.~A.} 2004 {\em {Turbulence : an introduction for scientists
  and engineers}\/}. Oxford, UK: Oxford University Press, 2004.

\bibitem[{Dormy} {\em et~al.\/}(2000){Dormy}, {Valet} \& {Courtillot}]{dormy00}
{\sc {Dormy}, E., {Valet}, J.-P. \& {Courtillot}, V.} 2000 {Numerical models of
  the geodynamo and observational constraints}. {\em Geochem. Geophys.
  Geosys.\/} {\bf 1}, 1037--42.

\bibitem[{Dyachenko} {\em et~al.\/}(1992){Dyachenko}, {Newell}, {Pushkarev} \&
  {Zakharov}]{Dyachenko}
{\sc {Dyachenko}, S., {Newell}, A.~C., {Pushkarev}, A. \& {Zakharov}, V.~E.}
  1992 {Optical turbulence: weak turbulence, condensates and collapsing
  filaments in the nonlinear Schr{\"o}dinger equation}. {\em Physica D\/} {\bf
  57}, 96--160.

\bibitem[{Falcon} {\em et~al.\/}(2007){Falcon}, {Laroche} \& {Fauve}]{Falcon}
{\sc {Falcon}, {\'E}., {Laroche}, C. \& {Fauve}, S.} 2007 {Observation of
  Gravity-Capillary Wave Turbulence}. {\em Phys. Rev. Lett.\/} {\bf 98}~(9),
  094503.

\bibitem[{Favier} {\em et~al.\/}(2012){Favier}, {Godeferd} \& {Cambon}]{favier}
{\sc {Favier}, B.F.N., {Godeferd}, F.S. \& {Cambon}, C.} 2012 {On the effect of
  rotation on magnetohydrodynamic turbulence at high magnetic Reynolds number}.
  {\em Geophys. Astrophys. Fluid Dyn.\/} {\bf 106}, 89--111.

\bibitem[{Finlay}(2008)]{Finlay}
{\sc {Finlay}, C.C.} 2008 Waves in the presence of magnetic fields, rotation
  and convection. In {\em Dynamos\/} (ed. P.~Cardin \& Elsevier
  science~publishers L.F. Cugliandolo~eds), {\em Les Houches 2007\/}, vol.~88,
  pp. 403--450.

\bibitem[{Finlay} {\em et~al.\/}(2010){Finlay}, {Dumberry}, {Chulliat} \&
  {Pais}]{Finlay10}
{\sc {Finlay}, C.~C., {Dumberry}, M., {Chulliat}, A. \& {Pais}, M.~A.} 2010
  {Short Timescale Core Dynamics: Theory and Observations}. {\em Space Sci.
  Rev.\/} {\bf 155}, 177--218.

\bibitem[{Finlay} \& {Jackson}(2003)]{finlay03}
{\sc {Finlay}, C.~C. \& {Jackson}, A.} 2003 {Equatorially Dominated Magnetic
  Field Change at the Surface of Earth's Core}. {\em Science\/} {\bf 300},
  2084--2086.

\bibitem[{Galtier}(2003)]{galtier03a}
{\sc {Galtier}, S.} 2003 {Weak inertial-wave turbulence theory}. {\em Phys.
  Rev. E\/} {\bf 68}~(1), 015301.

\bibitem[{Galtier}(2006{\natexlab{{\em a\/}}})]{galtier06b}
{\sc {Galtier}, S.} 2006{\natexlab{{\em a\/}}} {Multi-scale Turbulence in the
  Inner Solar Wind}. {\em J. Low Temperature Physics\/} {\bf 145}, 59--74.

\bibitem[{Galtier}(2006{\natexlab{{\em b\/}}})]{galtier06}
{\sc {Galtier}, S.} 2006{\natexlab{{\em b\/}}} {Wave turbulence in
  incompressible Hall magnetohydrodynamics}. {\em J. Plasma Phys.\/} {\bf 72},
  721--769.

\bibitem[{Galtier}(2009{\natexlab{{\em a\/}}})]{galtier09r}
{\sc {Galtier}, S.} 2009{\natexlab{{\em a\/}}} {Exact vectorial law for
  homogeneous rotating turbulence}. {\em Phys. Rev. E\/} {\bf 80}, 046301.

\bibitem[{Galtier}(2009{\natexlab{{\em b\/}}})]{Galtier09}
{\sc {Galtier}, S.} 2009{\natexlab{{\em b\/}}} {Wave turbulence in magnetized
  plasmas}. {\em Nonlin. Proc. Geophys.\/} {\bf 16}, 83--98.

\bibitem[{Galtier}(2014)]{galtier2014}
{\sc {Galtier}, S.} 2014 {Theory for helical turbulence under fast rotation}.
  {\em Phys. Rev. E\/} {\bf in press}.

\bibitem[{Galtier} \& {Bhattacharjee}(2003)]{galtier03b}
{\sc {Galtier}, S. \& {Bhattacharjee}, A.} 2003 {Anisotropic weak whistler wave
  turbulence in electron magnetohydrodynamics}. {\em Phys. Plasmas\/} {\bf 10},
  3065--3076.

\bibitem[{Galtier} \& {Chandran}(2006)]{galtierchandran}
{\sc {Galtier}, S. \& {Chandran}, B.~D.~G.} 2006 {Extended spectral scaling
  laws for shear-Alfv{\'e}n wave turbulence}. {\em Phys. Plasmas\/} {\bf
  13}~(11), 114505.

\bibitem[{Galtier} \& {Nazarenko}(2008)]{galtierNaza}
{\sc {Galtier}, S. \& {Nazarenko}, S.~V.} 2008 {Large-scale magnetic field
  sustainment by forced MHD wave turbulence}. {\em J. Turbulence\/} {\bf 9},
  40.

\bibitem[{Galtier} {\em et~al.\/}(2000){Galtier}, {Nazarenko}, {Newell} \&
  {Pouquet}]{galtier00}
{\sc {Galtier}, S., {Nazarenko}, S.~V., {Newell}, A.~C. \& {Pouquet}, A.} 2000
  {A weak turbulence theory for incompressible magnetohydrodynamics}. {\em J.
  Plasma Physics\/} {\bf 63}, 447--488.

\bibitem[{Galtier} {\em et~al.\/}(2002){Galtier}, {Nazarenko}, {Newell} \&
  {Pouquet}]{galtier02}
{\sc {Galtier}, S., {Nazarenko}, S.~V., {Newell}, A.~C. \& {Pouquet}, A.} 2002
  {Anisotropic Turbulence of Shear-Alfv{\'e}n Waves}. {\em Astrophys. J.
  Lett.\/} {\bf 564}, L49--L52.

\bibitem[{Glatzmaier} \& {Roberts}(1995)]{Glatzmaier}
{\sc {Glatzmaier}, G.~A. \& {Roberts}, P.~H.} 1995 {A three-dimensional
  self-consistent computer simulation of a geomagnetic field reversal}. {\em
  Nature\/} {\bf 377}, 203--209.

\bibitem[{Greenspan}(1968)]{greenspan}
{\sc {Greenspan}, H.P.} 1968 {\em {The Theory of Rotating Fluids}\/}. Cambridge
  University Press, 1968.

\bibitem[{Hasselmann}(1962)]{hasselman62}
{\sc {Hasselmann}, K.} 1962 {On the non-linear energy transfer in a
  gravity-wave spectrum. Part 1. General theory}. {\em J. Fluid Mech.\/} {\bf
  12}, 481--500.

\bibitem[{Hopfinger} {\em et~al.\/}(1982){Hopfinger}, {Gagne} \&
  {Browand}]{hop}
{\sc {Hopfinger}, E.~J., {Gagne}, Y. \& {Browand}, F.~K.} 1982 {Turbulence and
  waves in a rotating tank}. {\em J. Fluid Mech.\/} {\bf 125}, 505--534.

\bibitem[{Iroshnikov}(1964)]{iroshnikov64}
{\sc {Iroshnikov}, P.~S.} 1964 {Turbulence of a Conducting Fluid in a Strong
  Magnetic Field}. {\em Soviet Astron.\/} {\bf 7}, 566--571.

\bibitem[{Jacquin} {\em et~al.\/}(1990){Jacquin}, {Leuchter}, {Cambon} \&
  {Mathieu}]{jacquin}
{\sc {Jacquin}, L., {Leuchter}, O., {Cambon}, C. \& {Mathieu}, J.} 1990
  {Homogeneous turbulence in the presence of rotation}. {\em J. Fluid Mech.\/}
  {\bf 220}, 1--52.

\bibitem[{Jones}(2011)]{jones}
{\sc {Jones}, C.~A.} 2011 {Planetary Magnetic Fields and Fluid Dynamos}. {\em
  Ann. Rev. Fluid Mech.\/} {\bf 43}, 583--614.

\bibitem[{Kingsep} {\em et~al.\/}(1990){Kingsep}, {Chukbar} \&
  {Yankov}]{Kingsep}
{\sc {Kingsep}, A.S., {Chukbar}, K.V. \& {Yankov}, V.V.} 1990 Electron
  magnetohydrodynamics. In {\em Reviews of Plasma Physics\/} (ed. B.B.
  Kadomtsev), {\em Consultant Bureau, New York\/}, vol.~16, pp. 243--291.

\bibitem[{Kolmakov} {\em et~al.\/}(2004){Kolmakov}, {Levchenko}, {Brazhnikov},
  {Mezhov-Deglin}, {Silchenko} \& {McClintock}]{Kolmakov}
{\sc {Kolmakov}, G.~V., {Levchenko}, A.~A., {Brazhnikov}, M.~Y.,
  {Mezhov-Deglin}, L.~P., {Silchenko}, A.~N. \& {McClintock}, P.~V.} 2004
  {Quasiadiabatic Decay of Capillary Turbulence on the Charged Surface of
  Liquid Hydrogen}. {\em Phys. Rev. Lett.\/} {\bf 93}~(7), 074501.

\bibitem[Kraichnan(1965)]{kraichnan65}
{\sc Kraichnan, R.~H.} 1965 Inertial range spectrum in hydromagnetic
  turbulence. {\em Phys. Fluids\/} {\bf 8}, 1385--1387.

\bibitem[{Kraichnan}(1973)]{Kraichnan73}
{\sc {Kraichnan}, R.~H.} 1973 {Helical turbulence and absolute equilibrium}.
  {\em J. Fluid Mech.\/} {\bf 59}, 745--752.

\bibitem[{Kuznetsov}(1972)]{kuznetsov72}
{\sc {Kuznetsov}, E.~A.} 1972 {Turbulence of ion sound in a plasma located in a
  magnetic field}. {\em Sov. Phys. J. Exp. Theor. Phys.\/} {\bf 35}, 310--314.

\bibitem[{Lamriben} {\em et~al.\/}(2011){Lamriben}, {Cortet} \&
  {Moisy}]{lamriben}
{\sc {Lamriben}, C., {Cortet}, P.-P. \& {Moisy}, F.} 2011 {Direct Measurements
  of Anisotropic Energy Transfers in a Rotating Turbulence Experiment}. {\em
  Phys. Rev. Lett.\/} {\bf 107}~(2), 024503.

\bibitem[{Lehnert}(1954)]{lehnert}
{\sc {Lehnert}, B.} 1954 {Magnetohydrodynamic Waves Under the Action of the
  Coriolis Force.} {\em Astrophys. J.\/} {\bf 119}, 647.

\bibitem[Lesieur(1997)]{lesieur97}
{\sc Lesieur, M.} 1997 {\em Turbulence in Fluids\/}. 3rd ed. Kluwer Academic.

\bibitem[{Lvov} {\em et~al.\/}(2003){Lvov}, {Nazarenko} \& {West}]{lvov03}
{\sc {Lvov}, Y., {Nazarenko}, S. \& {West}, R.} 2003 {Wave turbulence in
  Bose-Einstein condensates}. {\em Physica D\/} {\bf 184}, 333--351.

\bibitem[{Matthaeus} \& {Goldstein}(1982)]{matthaeus82}
{\sc {Matthaeus}, W.~H. \& {Goldstein}, M.~L.} 1982 {Measurement of the rugged
  invariants of magnetohydrodynamic turbulence in the solar wind}. {\em J.
  Geophys. Res.\/} {\bf 87}, 6011--6028.

\bibitem[Meyrand \& Galtier(2012)]{Meyrand12}
{\sc Meyrand, R. \& Galtier, S.} 2012 Spontaneous chiral symmetry breaking of
  hall mhd turbulence. {\em Phys. Rev. Lett.\/} {\bf 109}, 194501.

\bibitem[{Mininni} \& {Pouquet}(2009)]{mininni09}
{\sc {Mininni}, P.~D. \& {Pouquet}, A.} 2009 {Helicity cascades in rotating
  turbulence}. {\em Phys. Rev. E\/} {\bf 79}~(2), 026304.

\bibitem[{Mininni} \& {Pouquet}(2010{\natexlab{{\em a\/}}})]{mininni10a}
{\sc {Mininni}, P.~D. \& {Pouquet}, A.} 2010{\natexlab{{\em a\/}}} {Rotating
  helical turbulence. I. Global evolution and spectral behavior}. {\em Phys.
  Fluids\/} {\bf 22}~(3), 035105.

\bibitem[{Mininni} \& {Pouquet}(2010{\natexlab{{\em b\/}}})]{mininni10b}
{\sc {Mininni}, P.~D. \& {Pouquet}, A.} 2010{\natexlab{{\em b\/}}} {Rotating
  helical turbulence. II. Intermittency, scale invariance, and structures}.
  {\em Phys. Fluids\/} {\bf 22}~(3), 035106.

\bibitem[{Mininni} {\em et~al.\/}(2012){Mininni}, {Rosenberg} \&
  {Pouquet}]{mininni12}
{\sc {Mininni}, P.~D., {Rosenberg}, D. \& {Pouquet}, A.} 2012 {Isotropization
  at small scales of rotating helically driven turbulence}. {\em J. Fluid
  Mech.\/} {\bf 699}, 263--279.

\bibitem[{Moffatt}(1969)]{moffatt}
{\sc {Moffatt}, H.~K.} 1969 {The degree of knottedness of tangled vortex
  lines}. {\em J. Fluid Mech.\/} {\bf 35}, 117--129.

\bibitem[{Moffatt}(1970)]{moffatt70}
{\sc {Moffatt}, H.~K.} 1970 {Dynamo action associated with random inertial
  waves in a rotating conducting fluid}. {\em J. Fluid Mech.\/} {\bf 44},
  705--719.

\bibitem[{Moffatt}(1972)]{moffatt72}
{\sc {Moffatt}, H.~K.} 1972 {An approach to a dynamic theory of dynamo action
  in a rotating conducting fluid}. {\em J. Fluid Mech.\/} {\bf 53}, 385--399.

\bibitem[{Moffatt}(1978)]{moffatt78}
{\sc {Moffatt}, H.~K.} 1978 {\em {Magnetic field generation in electrically
  conducting fluids}\/}. Cambridge, England, Cambridge University Press, 1978.

\bibitem[{Morin} {\em et~al.\/}(2011){Morin}, {Dormy}, {Schrinner} \&
  {Donati}]{morin}
{\sc {Morin}, J., {Dormy}, E., {Schrinner}, M. \& {Donati}, J.-F.} 2011 {Weak-
  and strong-field dynamos: from the Earth to the stars}. {\em Mon. Not. R.
  Astron. Soc.\/} {\bf 418}, L133--L137.

\bibitem[{Morize} {\em et~al.\/}(2005){Morize}, {Moisy} \& {Rabaud}]{morize}
{\sc {Morize}, C., {Moisy}, F. \& {Rabaud}, M.} 2005 {Decaying grid-generated
  turbulence in a rotating tank}. {\em Phys. Fluids\/} {\bf 17}~(9), 095105.

\bibitem[Nazarenko(2011)]{nazarenko11}
{\sc Nazarenko, S.} 2011 {\em Wave Turbulence\/}. Lecture Notes in Physics,
  Berlin Springer Verlag.

\bibitem[{Newell} {\em et~al.\/}(2001){Newell}, {Nazarenko} \&
  {Biven}]{newell01}
{\sc {Newell}, A.~C., {Nazarenko}, S. \& {Biven}, L.} 2001 {Wave turbulence and
  intermittency}. {\em Physica D Nonlinear Phenomena\/} {\bf 152}, 520--550.

\bibitem[{Perez} \& {Boldyrev}(2008)]{perez08}
{\sc {Perez}, J.~C. \& {Boldyrev}, S.} 2008 {On Weak and Strong
  Magnetohydrodynamic Turbulence}. {\em Astrophys. J. Lett.\/} {\bf 672},
  L61--L64.

\bibitem[{P{\'e}tr{\'e}lis} {\em et~al.\/}(2007){P{\'e}tr{\'e}lis}, {Mordant}
  \& {Fauve}]{ens}
{\sc {P{\'e}tr{\'e}lis}, F., {Mordant}, N. \& {Fauve}, S.} 2007 {On the
  magnetic fields generated by experimental dynamos}. {\em Geophys. Astrophys.
  Fluid Dyn.\/} {\bf 101}, 289--323.

\bibitem[{Pouquet} {\em et~al.\/}(1976){Pouquet}, {Frisch} \&
  {Leorat}]{pouquet76}
{\sc {Pouquet}, A., {Frisch}, U. \& {Leorat}, J.} 1976 {Strong MHD helical
  turbulence and the nonlinear dynamo effect}. {\em J. Fluid Mech.\/} {\bf 77},
  321--354.

\bibitem[{Roberts} \& {King}(2013)]{Roberts}
{\sc {Roberts}, P.~H. \& {King}, E.~M.} 2013 {On the genesis of the Earth's
  magnetism}. {\em Reports Prog. Physics\/} {\bf 76}~(9), 096801.

\bibitem[{Sagdeev} \& {Galeev}(1969)]{sagdeev69}
{\sc {Sagdeev}, R.~Z. \& {Galeev}, A.~A.} 1969 {\em {Nonlinear Plasma
  Theory}\/}. Nonlinear Plasma Theory, New York: Benjamin, 1969.

\bibitem[{Sahraoui} {\em et~al.\/}(2007){Sahraoui}, {Galtier} \&
  {Belmont}]{sahraoui07}
{\sc {Sahraoui}, F., {Galtier}, S. \& {Belmont}, G.} 2007 {On waves in
  incompressible Hall magnetohydrodynamics}. {\em J. Plasma Phys.\/} {\bf 73},
  723--730.

\bibitem[{Schmitt} {\em et~al.\/}(2008){Schmitt}, {Alboussi{\`e}re}, {Brito},
  {Cardin}, {Gagni{\`e}re}, {Jault} \& {Nataf}]{schmitt08}
{\sc {Schmitt}, D., {Alboussi{\`e}re}, T., {Brito}, D., {Cardin}, P.,
  {Gagni{\`e}re}, N., {Jault}, D. \& {Nataf}, H.-C.} 2008 {Rotating spherical
  Couette flow in a dipolar magnetic field: experimental study of
  magneto-inertial waves}. {\em J. Fluid Mech.\/} {\bf 604}, 175--197.

\bibitem[{Scott}(2014)]{scott}
{\sc {Scott}, J.~F.} 2014 {Wave turbulence in a rotating channel}. {\em J.
  Fluid Mech.\/} {\bf 741}, 316--349.

\bibitem[{Shaikh} \& {Zank}(2005)]{shaikh05}
{\sc {Shaikh}, D. \& {Zank}, G.~P.} 2005 {Driven dissipative whistler wave
  turbulence}. {\em Phys. Plasmas\/} {\bf 12}~(12), 122310.

\bibitem[{Shebalin}(2006)]{shebalin06}
{\sc {Shebalin}, J.~V.} 2006 {Ideal homogeneous magnetohydrodynamic turbulence
  in the presence of rotation and a mean magnetic field}. {\em J. Plasma
  Phys.\/} {\bf 72}, 507--524.

\bibitem[{Shirley} \& {Fairbridge}(1997)]{shirley}
{\sc {Shirley}, J.~H. \& {Fairbridge}, R.~W.} 1997 {\em {Encyclopedia of
  Planetary Sciences}\/}. Springer-Verlag, Berlin Heidelberg, 1997.

\bibitem[{Smith} \& {Lee}(2005)]{smith2}
{\sc {Smith}, L.~M. \& {Lee}, Y.} 2005 {On near resonances and symmetry
  breaking in forced rotating flows at moderate Rossby number}. {\em J. Fluid
  Mech.\/} {\bf 535}, 111--142.

\bibitem[{Smith} \& {Waleffe}(1999)]{smith1}
{\sc {Smith}, L.~M. \& {Waleffe}, F.} 1999 {Transfer of energy to
  two-dimensional large scales in forced, rotating three-dimensional
  turbulence}. {\em Phys. Fluids\/} {\bf 11}, 1608--1622.

\bibitem[{Stevenson}(2003)]{stevenson}
{\sc {Stevenson}, D.~J.} 2003 {Planetary magnetic fields}. {\em Earth Planet.
  Sci. Lett.\/} {\bf 208}, 1--11.

\bibitem[{Teitelbaum} \& {Mininni}(2009)]{teitelbaum09}
{\sc {Teitelbaum}, T. \& {Mininni}, P.~D.} 2009 {Effect of Helicity and
  Rotation on the Free Decay of Turbulent Flows}. {\em Phys. Rev. Lett.\/} {\bf
  103}~(1), 014501.

\bibitem[{Turner}(2000)]{turner00}
{\sc {Turner}, L.} 2000 {Using helicity to characterize homogeneous and
  inhomogeneous turbulent dynamics}. {\em J. Fluid Mech.\/} {\bf 408},
  205--238.

\bibitem[{van Bokhoven} {\em et~al.\/}(2009){van Bokhoven}, {Clercx}, {van
  Heijst} \& {Trieling}]{vanbokhoven}
{\sc {van Bokhoven}, L.~J.~A., {Clercx}, H.~J.~H., {van Heijst}, G.~J.~F. \&
  {Trieling}, R.~R.} 2009 {Experiments on rapidly rotating turbulent flows}.
  {\em Phys. Fluids\/} {\bf 21}~(9), 096601.

\bibitem[{Waleffe}(1992)]{waleffe92}
{\sc {Waleffe}, F.} 1992 {The nature of triad interactions in homogeneous
  turbulence}. {\em Phys. Fluids\/} {\bf 4}, 350--363.

\bibitem[{Waleffe}(1993)]{waleffe2}
{\sc {Waleffe}, F.} 1993 {Inertial transfers in the helical decomposition}.
  {\em Phys. Fluids\/} {\bf 5}, 677--685.

\bibitem[{Zakharov}(1965)]{Zakh65}
{\sc {Zakharov}, V.~E.} 1965 {Weak turbulence in media with a decay spectrum}.
  {\em J. Appl. Mech. Tech. Phys.\/} {\bf 6}, 22--24.

\bibitem[{Zakharov}(1967)]{zakharov67}
{\sc {Zakharov}, V.~E.} 1967 {On the spectrum of turbulence in plasma without
  magnetic field}. {\em J. Exp. Theor. Phys.\/} {\bf 24}, 455--459.

\bibitem[{Zakharov} \& Filonenko(1966)]{Zakh66}
{\sc {Zakharov}, V.~E. \& Filonenko, N.N.} 1966 {The energy spectrum for
  stochastic oscillations of a fluid surface}. {\em Doclady Akad. Nauk. SSSR\/}
  {\bf 170}, 1292--1295.

\bibitem[{Zakharov} {\em et~al.\/}(1992){Zakharov}, {L'Vov} \&
  {Falkovich}]{ZLF92}
{\sc {Zakharov}, V.~E., {L'Vov}, V.~S. \& {Falkovich}, G.} 1992 {\em
  {Kolmogorov spectra of turbulence I: Wave turbulence}\/}. Springer Series in
  Nonlinear Dynamics, Berlin: Springer, 1992.

\end{thebibliography}
\end{document}